\documentstyle[12pt]{article}
\newcommand{\half}{{\scriptstyle{{1\over 2}}}}
\newcommand{\quarter}{{\scriptstyle{{1\over 4}}}}
\def\beq{\begin{equation}}
\def\eeq{\end{equation}}
\def\bea{\begin{array}}
\def\eea{\end{array}}
\def\beqa{\begin{eqnarray}}
\def\eeqa{\end{eqnarray}}
\newcommand{\refeq}[1]{\mbox{eq.~(\ref{eq:#1})}}
\def\asd{an\-ti-self\-du\-al}
\def\sd{self-du\-al}
\def\myIm{{\Im m}}
\def\myre{{\rm Re}}
\def\myRe{{\Re e}}
\def\myim{{\rm Im}}
\def\etabar{{\bar{\eta}}}
\def\u1{{U(1)}}
\def\su2{{SU(2)}}

\def\FT{Fourier~trans\-for\-ma\-tion}
\def\ie{{i.e.\/}}\relax
\def\eg{{e.g.\/}}\relax
\def\cf{{cf.\/}}\relax
\newcommand{\zahlen}{{\rm Z \!\! Z}}
\newcommand{\re}{\relax{\rm I\kern-.18em R}}
\newcommand{\qu}{{\rm I \! H}}
\def\bsigma{{\bar\sigma}}
\def\cT{{\cal{T}}}
\def\cO{{\cal{O}}}
\def\cP{{\cal{P}}}
\def\cM{{\cal{M}}}
\def\cY{\hat{\cal{Y}}}
\def\cC{\hat{\cal{C}}}
\def\ddz{{\frac{d}{dz}}}
\def\Tr{{\rm Tr}} 
\def\tr{{\rm tr}}

\begin{document}
\hfill INLO-PUB-5/98
\vskip1cm
\begin{center}
{\LARGE{\bf{\underline{Periodic Instantons with}}}}\\
{\LARGE{\bf{\underline{non-trivial Holonomy}}}}\\
\vspace{7mm}
{\large Thomas C. Kraan and Pierre van Baal} \\
\vspace{5mm}
Instituut-Lorentz for Theoretical Physics, University of Leiden,\\
PO Box 9506, NL-2300 RA Leiden, The Netherlands.
\end{center}
\vspace*{5mm}{\narrower\narrower{\noindent
\underline{Abstract:} 
We present the detailed derivation of the charge one periodic instantons - or
calorons - with non-trivial holonomy for $\su2$. We use a suitable combination
of the Nahm transformation and ADHM techniques. Our results rely on our ability
to compute explicitly the relevant Green's function in terms of which the 
solution can be conveniently expressed. We also discuss the properties of the 
moduli space, $\re^3 \times S^1 \times$ Taub-NUT/$Z_2$ and its metric, relating
the holonomy to the Taub-NUT mass parameter. We comment on the monopole 
constituent description of these calorons, how to retrieve topological charge 
in the context of abelian projection and possible applications to QCD.}\par}
\section{Introduction}
Instantons~\cite{AtiDriHitMan} and Bogomolny-Prasad-Sommerfield (BPS) 
monopoles~\cite{BPS} possess remarkable properties. They exist as exact 
solutions with arbitrary charges and with an action or energy, proportional to 
their integer charge. Therefore the multi-charge solutions can be seen as built
from constituents of unit charge. Indeed, for BPS monopoles the absence of an 
interaction energy can be understood - for large separation - as a cancellation
between the electro-magnetic and scalar interactions~\cite{Man0,MoOl}. 

Instantons are \sd\ solutions with finite action. For non-compact manifolds, 
the solutions must approach vacua in the non-compact directions. Due to the 
topology of the base manifold, these vacua can be non-trivial and can give 
rise to extra parameters for these \sd\ solutions. For periodic instantons 
on $\re^3 \times S^1$, also called calorons, the vacuum label is given by the 
eigenvalues of the Polyakov loop (holonomy) around $S^1$ at spatial 
infinity. Equivalently, one may consider this vacuum as the background 
field on which the solution is superposed. In this respect, they are very 
similar to monopole solutions in broken gauge theories, a non-trivial vacuum 
generally breaking the gauge symmetry. 

Calorons can be seen to have as constituents BPS monopoles~\cite{GarMur,LeYi} ($
N$ for $SU(N)$), as follows from Nahm's work~\cite{NahMonCal}. The constituents 
are such that the net magnetic and electric charge of the caloron vanishes. 
Unlike for the ordinary multi-monopoles, the BPS constituents are hence of 
opposite charge, and thus have an attractive electro-magnetic interaction. 
Nevertheless, also here exact solutions exist with an {\em action} that does 
not depend on the parameters, though the solutions become static only for large
separations. In order to have this non-trivial situation the Polyakov loop 
at spatial infinity has to be non-trivial, breaking the gauge invariance 
spontaneously. The eigenvalues of this Polyakov loop uniquely fix the masses 
of the constituent monopoles. Their separation - not surprisingly - 
is related to the scale parameter of the caloron solution.

In this paper we study periodic $\su2$ instantons with topological charge one 
and arbitrary holonomy~\cite{PLB}. Its purpose is to provide the necessary 
details for this construction. Central to our success in providing explicit 
and relatively simple new solutions is the construction of the relevant Green's
function. In the context of the Nahm transformation, introduced here as the 
Fourier transform of the Atiyah-Drinfeld-Hitchin-Manin (ADHM) 
data~\cite{AtiDriHitMan}, this can be reduced to a quantum mechanical problem 
on the circle with a piecewise constant potential and well-defined delta 
function singularities related to the holonomy. We find compact expressions 
for the gauge field and action density of the solution and investigate the 
properties of the caloron. The moduli space is described in terms of the 
constituent monopoles, which in our approach appear as explicit lumps in the 
action density. Furthermore, we relate the constituent monopole nature of 
these instantons to work by Taubes in which he showed how to make gauge 
configurations with non-trivial topological charge out of monopole 
fields~\cite{Taubes}. 

Independently, the recent work in ref.~\cite{LeeLu} has taken the constituent 
monopole description~\cite{GarMur,LeYi,NahMonCal} as the starting point, 
suitably superposing two BPS monopole solutions to form a caloron solution. 

Periodic instantons have been discussed first in the context of finite 
temperature field theory~\cite{HarShe,GroPisYaf}, where the period ($\cT$) is 
the inverse temperature in euclidean field theory. A non-trivial value of the 
Polyakov loop will modify the vacuum fluctuations and thereby leads to a 
non-zero vacuum energy density as compared to a trivial Polyakov loop. It was 
on the basis of this observation that calorons with non-trivial holonomy were 
deemed irrelevant in the infinite volume limit~\cite{GroPisYaf}. It should be
emphasised though, that the semi-classical one-instanton calculation is no 
longer considered a reliable approximation. At finite temperature $A_0$ can 
be seen to play the role of a Higgs field and in a strongly interacting 
environment one could envisage regions with this Higgs field 
pointing predominantly in a certain direction, and nevertheless having at 
infinity a trivial Higgs field. Given a finite density of periodic instantons, 
in an infinite volume solutions with non-trivial holonomy (in some average 
sense) may well have a role to play.

In our construction of the charge one caloron with non-trivial Polyakov loop,
we pick a particular gauge. In the periodic gauge, the spatial components of
the vacuum connection at infinity can be gauged to zero. The $A_0$ component
can only be gauged to a constant, \eg $A_0=2\pi i{\vec\omega}\cdot{\vec\tau}$
(with $\tau_a$ the Pauli matrices), when the total magnetic charge is vanishing.
This connection has obviously a non-trivial Polyakov loop at infinity
\beq
\cP(\vec x)=P\,\exp(\int_0^\cT A_0(\vec x,x_0)dx_0)\rightarrow 
e^{2\pi i\vec\omega\cdot\vec\tau}\label{eq:polloop}
\eeq
($P$ stands for path-ordering).
Alternatively, connections on $\re^3 \times S^1$ can be formulated by 
embedding them in $\re^4$ and demanding periodicity modulo gauge 
transformations. Gauging with a non-periodic gauge transformation 
$g(\vec x, x_0)=e^{2\pi ix_0\vec\omega\cdot\vec\tau}$, starting from the 
periodic gauge with zero $A_i$ and constant $A_0=2\pi i\vec\omega\cdot
\vec\tau$ at infinity, sets {\em all} gauge fields to zero at infinity.
In that case we have
\beq
A_\mu(\vec x,x_0+\cT)=e^{2\pi i\vec\omega\cdot\vec\tau}
A_\mu(\vec x, x_0)e^{-2\pi i\vec\omega\cdot\vec\tau},
\label{eq:permodgau}
\eeq
with $\cT$ the period in the imaginary time direction, i.e. the inverse 
temperature. Clearly, $e^{2\pi i\vec\omega\cdot\vec\tau}\equiv g_0({\vec x})$
is the transition function or cocycle. Using the proper expression for the
Polyakov loop along a path traversing the boundary between coordinate 
patches~\cite{BPr},
\beq
\cP(\vec x)=P\,\exp(\int_0^\cT A_0(\vec x,x_0)dx_0)g_0(\vec x),
\eeq
we find the same value for the holonomy in this 
gauge, the holonomy at infinity now solely being carried by the cocycle 
$g_0({\vec x})$. It is in this so-called ``algebraic" gauge that we will 
calculate the generalised caloron solutions.

The charge $k$ instantons on $\re^4$ are given by the ADHM
construction~\cite{AtiDriHitMan}. The Nahm
transformation forms a modification of this approach, initially introduced by 
Nahm to study BPS monopoles~\cite{NahFou,NahAll}. Later developments culminated
in the Nahm duality transformation on generalised tori, which forms a powerful 
tool for studying \sd\ connections. Also caloron solutions can be treated along
these lines~\cite{NahMonCal}. The Nahm transformation is outlined in section 2.
We summarise in section 3 the details of the ADHM formalism necessary for our 
construction of the caloron in section 4. We make it evident how the two 
approaches are related by Fourier transformation. By relying on the ADHM 
construction we profit from the vast knowledge on multi-instanton calculus 
within this formalism. Section 4 forms the calculational core of this paper, 
in which we derive the gauge potential and a particularly simple expression
for the action density. The various properties and symmetries of the caloron 
are unraveled in section 5. In section 6 we describe the moduli space of the 
caloron. In section 7 we discuss the relation to Taubes's work, abelian 
projection~\cite{Abproj} and possible applications to QCD.

\section{The Nahm transformation}
We will consider a $U(n)$ bundle $E$ with \sd\ gauge connection $A_\mu$ on a 
four manifold $M=\re^4/H$, with instanton number $k$. Here $H$ is a subgroup 
of translation symmetries under which the physics is invariant. When $H$ is a 
four dimensional lattice, $M$ will be the four torus~\cite{BraBaa}. Other four 
manifolds are obtained by taking appropriate limits~\cite{BaaT3R}. We demand 
the gauge potential to be invariant modulo gauge transformations under the 
action of $H$.

An essential ingredient in Nahm's construction~\cite{NahAll} is to add a 
curvature free abelian connection, $-2\pi iz_\mu dx_\mu$, to the gauge
field and to study the Weyl operator
\beq
D_z(A)=\sigma_\mu D^\mu_z(A),\quad D^\dagger_z(A)=-\bsigma_\mu D_z^\mu(A),\quad
D^\mu_z(A)=\partial_\mu+A_\mu-2\pi iz_\mu.\label{eq:modwey}
\eeq
$\sigma_\mu=(1_2,i\vec\tau)$ and $\bsigma_\mu=\sigma^\dagger_\mu=(1_2,-i\vec
\tau)$ are unit quaternions. As compared to usual 
conventions~\cite{BraBaa,BaaT3R}, we replaced $z$ by $-z$ to facilitate 
matching with the ADHM construction. When $A$ is without flat factors (WFF, 
meaning that the vector bundle $E$ does not split in $E'\oplus L$ for any 
flat line bundle $L$), then $D_z(A)$ will have a trivial kernel~\cite{DonKro}. 
For such gauge fields $G_z(x,y)=(D^\dagger_z(A)D_z(A))^{-1}$ is well-defined. 
The index theorem~\cite{AtiSinInd,CalBotSee} shows that there are $k$ 
normalisable zero modes of the Weyl operator $D^\dagger_z(A)$, for each value 
of $z\in\hat{M}=\re^4/\hat{H}$, $\hat{H}=\{z\in\re^4|z\cdot y\in\zahlen,\,
\forall y\in H\}$, \cf\ ref.~\cite{BaaT3R}. We can therefore define a $U(k)$ 
connection on the space $\hat M$,
\beq
\hat{A}^{ij}_\mu(z)=\int_Mdx~\psi_z^{i\dagger}(x)\frac{\partial}{\partial z_\mu}
\psi_z^j(x),
\eeq
where $\psi^i_z(x),~i=1,\ldots,k$ form an orthonormal basis for the Nahm 
bundle $\hat E$ of fermionic zero modes. This is called the Nahm transformed 
connection.

The Weitzenb\"ock identity~\cite{DonKro} states
\beq
D^\dagger_z(A)D_z(A)=-(D_z^\mu(A)D_z^\mu(A)+\half\bar{\eta}_{\mu\nu}
F^{\mu\nu}(x)).\label{eq:Weitzenboeck}
\eeq
where $\bar\eta_{\mu\nu}=\bar\eta^a_{\mu\nu}\sigma_a=\bsigma_{[\mu}
\sigma_{\nu]}$ is the \asd\ and $\eta_{\mu\nu}=\eta^a_{\mu\nu}\sigma_a=
\sigma_{[\mu}\bsigma_{\nu]}$ the \sd\  't Hooft tensor~\cite{Hoo} 
(note that in our conventions time is labeled by $x_0$ rather than $x_4$, and
to conform with \asd ity of $\bar\eta$ we define $\epsilon_{1230}=1$). As $F$ 
is \sd, the second term will vanish, and hence $D^\dagger_z(A)D_z(A)$ and 
$G_z(x,y)$ commute with the quaternions. This has profound consequences for 
the curvature associated to the Nahm connection. One finds~\cite{BaaT3R}
\begin{eqnarray}\label{eq:Nahmcurvature}
&&\hat{F}_{\mu\nu}(z)=8\pi^2\int_{M\times M}dxdy\,\psi_z^\dagger(x)G_z(x,y)
\eta_{\mu\nu}\psi_z(y)\\&&\qquad+4\pi i\int_{\partial M\times M}dS_\lambda(x)\,
dy\,\frac{\partial}{\partial z_{\left[\mu\right.}}\psi_z^\dagger(x)
\sigma_\lambda\bar\sigma_{\left.\nu\right]}G_z(x,y)\psi_z(y)\nonumber,
\end{eqnarray}
where $\psi_z$ denotes the matrix with the zero modes $\psi^i_z$ as columns.
Here we used that $G$ commutes with the quaternions. The first term is clearly 
\sd . The boundary term shows possible deviations from \sd ity, which occur
at the points $z$ for which the zero modes do not decay exponentially in the 
non-compact directions. In these directions the connection necessarily 
approaches a vacuum for the action to be finite. These vacua are labeled by 
the eigenvalues of the Polyakov loops $\cP_i=P\exp\int_{C_i}A_\mu dx_\mu$ 
along the circles $C_i$ corresponding to the compact directions. In the case 
that $e^{2\pi iz}$ becomes equal to one of these eigenvalues, the component 
of $A_\mu-2\pi iz_\mu$ along $C_i$ in \refeq{modwey} will develop a zero 
eigenvalue when approaching infinity. This gives rise to a surviving boundary 
term in \refeq{Nahmcurvature} and as a result a deviation from \sd ity, 
precisely for these specific points. As the deviations occur in single points, 
they will be expressible in delta functions. Hence, $\hat{F}_{\mu\nu}$ is 
\sd\ almost everywhere. For the non-compact directions $\mu$, the $z_\mu$ 
dependence of $\psi_z(x)$ is a plane wave factor, and hence $\hat{A}$ is 
$z_\mu$ independent. Note that the $U(k)$ symmetry in the space of zero 
modes associated to $A$ is mapped onto a gauge symmetry for $\hat{A}$. On 
the other hand, gauge transformations on $A$ leave $\hat{A}$ unchanged.

For the four torus $T^4$ the boundary terms are absent and instantons are
mapped onto instantons. It can be shown, using the family index theorem,
that under this Nahm transformation a $U(n)$ connection with topological 
charge $k$ is mapped onto a $U(k)$ connection with topological charge $n$.
The Nahm transformation on $T^4$ squares to the identity~\cite{BraBaa}. 
More explicitly, the dual Weyl operator $\hat D^\dagger_x(\hat A)=
-\bsigma_\mu(\hat\partial_\mu+\hat A_\mu-2\pi ix_\mu)$ has $n$ zero
modes
\beq
\hat D^\dagger_x(\hat A)\hat\psi_x^i(z)=0\label{eq:dualweyzm},\quad i=1,\ldots,n
\eeq
in terms of which the original connection $A_\mu(x)$ is reconstructed
\beq
A^{ij}_\mu(x)=\int_{\hat M}dz~\hat\psi^{\dagger i}_x(z)\frac{\partial}{
\partial x_\mu}\hat\psi_x^j(z).
\eeq
This suggests to use the Nahm transformation in the construction of \sd\
connections on modified tori, in situations when one can explicitly find the
dual connection $\hat A$.  Generally one expects this to be feasible when the
Nahm transformed bundle is simpler than the original, in particular when $\hat
M$ is of lower dimension than $M$. Another simplification arises for the
case of topological charge $k=1$, since in that case the Nahm connection
$\hat A$ is abelian. Because of the boundary terms, the second Nahm 
transformation will have to be modified by properly handling the singularities.
The extreme case is $M=\re^4,~H=0$, where the dual manifold $\hat{M}$ is just 
a point and the pair formed by the dual Weyl operator and singularities reduce 
to matrices which precisely give the ADHM data~\cite{NahMonCal,BaaT3R,DonKro,
CorGod}. The Nahm transformation on $\re^4/H$ encompasses the ADHM construction.

We will now consider the Nahm transformation for calorons (using the classical
scale invariance of the \sd ity equations, we can choose $\cT=1$ such that
$H = \zahlen$) and monopoles in the BPS limit ($H=\re$). For the latter, 
$A_0$ is interpreted as the Higgs field. Thus we can unify these two cases by
considering them as connections on $\re^3 \times S^1$. The connections for
$\re^3 \times S^1$ are topologically classified according to their behaviour at
the boundary, $S^2 \times S^1$. We give a short summary of the classification
presented in ref.~\cite{GroPisYaf}. 

For the action to be finite, it is necessary that the connections go to a 
vacuum at spatial infinity. Generally, gauge vacua are labeled by the 
conjugacy classes of representations of maps of the first homotopy group to 
the gauge group~\cite{DonKro}. For $S^2\times S^1$, we can characterise each 
vacuum by a gauge equivalence class of an element of the gauge group. Using 
a gauge transformation, this element can be chosen diagonal. The vacuum at 
infinity is related to the holonomy along the $S^1$ (or Polyakov loop), 
\refeq{polloop} in the periodic gauge. The difference of a closed Wilson loop 
evaluated along two curves $C$ and $C'$ is related with the flux through the 
surface swept out by the curves interpolating between $C$ and $C'$. Hence, 
at spatial infinity where the curvature vanishes, a small deformation of the 
path $C$ around which the holonomy is measured does not influence 
$\cP({\vec x})$. Only the homotopy of $C$ is important, and the holonomy 
at spatial infinity is a topological invariant. Therefore, at spatial infinity 
the eigenvalues of $\cP({\vec x})$ become constants
\beq
\cP({\vec x}) \rightarrow V(\hat x)\exp[2\pi i\,{\rm diag}(\mu_1,\cdots,\mu_n)]
V^{-1}(\hat x),\quad\sum_i\mu_i=0,\label{eq:rotdiag}
\eeq
and we have, up to an $\hat x$ dependent gauge transformation $V$,
\beq
A_0=2\pi i\,{\rm diag}(\mu_1,\cdots,\mu_n)-i\,{\rm diag}(k_1,\cdots,k_n)/2r+
\cO(r^{-2}),\quad\sum_i k_i=0.\label{eq:asa0}
\eeq
The gauge transformation $V$ induces a map from $S^2$ to the factor group 
$G/H_\infty$, with $H_\infty$ the isotropy group of $\exp[2\pi i\,{\rm diag}
(\mu_1,\cdots,\mu_n)]$. For $SU(n)$ these maps $V(\hat x)\rightarrow SU(n)
/H_\infty$ are classified according to the fundamental group of $H_\infty$. 
Generically, $H_\infty$ consists of several $U(1)$ and $SU(N),\,N >1$ 
subgroups. Each $U(1)$ gives rise to a monopole winding number, related to 
the integers $k_i$.

The other topological quantum number is related to the homotopy class
of the map $\partial M = S^2 \times S^1 \rightarrow SU(N)$ which occurs in the
gauge transformation connecting the behaviour near the origin to that at
infinity, which is classified by the instanton number $k\in\pi_3(SU(n))=
\zahlen$. Gauge connections on $\re^3 \times S^1$ are therefore classified
by $\mu_i$, $k_i$ and $k$. We will consider the situation where the net 
magnetic charges of the solution vanish, $k_i=0$. For non-zero $k_i$ the 
situation is as for the (static) BPS monopoles, where the dimension of the 
space of fermionic zero modes depends on $z$. The jumps occur exactly where 
$z=\mu_i$, according to the Callias-Bott-Seeley theorem~\cite{CalBotSee} 
(the situation of non-maximal symmetry breaking, where two or more $\mu_i$ 
coincide, is more involved).

For a periodic instanton with no net magnetic charges the fermionic zero modes 
are associated to the instanton winding number $k$ in the usual way. Hence, the 
rank of the Nahm transformed gauge potential is $k$ and as the ${\vec z}$ 
dependence of the fermionic zero modes is that of a plane wave,
$\psi_{z_0,\vec z}(x)=e^{2\pi i\vec x\cdot\vec z}\psi_{z_0,\vec 0}(x)$,
one finds $\hat{A}$ to be ${\vec z}$ independent. The dual space $\hat M$ is 
an interval on the real line, with coordinate $z_0\equiv z$. For monopoles, 
this interval is $[\mu_1,\mu_n]$ (when we order $\mu_i\leq\mu_{i+1}$). For 
calorons $z$ is the coordinate on the dual circle and $\hat A$ is periodic. 
The Nahm transformed curvature reduces to
\beq
\hat{F}_{0i}(z)=\ddz \hat{A}_i+[\hat{A}_0,\hat{A}_i],\quad
\hat{F}_{ij}(z)=[\hat{A}_i,\hat{A}_j].
\eeq
Using the \sd ity of the first term in \refeq{Nahmcurvature}, and the fact 
that the second term of this equation is zero almost everywhere, except for
possible delta function singularities at $z=\mu_i$ one finds 
\beq
\ddz\hat{A}_i+[\hat{A}_0,\hat{A}_i]+\half\epsilon_{ijk}[\hat{A}_j,\hat{A}_k]
=\sum_i\alpha_i\delta(z-\mu_i).
\eeq
These are the celebrated Nahm equations. Any $\hat A_\mu(z)$ obeying the Nahm 
equation of the right dimensionalities and singularity structure gives rise 
to a BPS monopole or caloron. For monopoles this is generally proven using 
twistor methods~\cite{NahAll,Hit}, but for $\su2$ monopoles there exist 
direct proofs of the equivalence, without an intermediate twistor 
step~\cite{Nak}. For calorons the construction was formulated in 
ref.~\cite{NahMonCal} and the twistor method for these periodic instantons 
was given in ref.~\cite{GarMur}, but a relation with existence theorems or 
a full circle reciprocity proof as it exists for monopoles and for instantons 
on $\re^4$ and $T^4$ seems not to be present.

For the $k=1$ $\su2$ caloron, $\hat A_\mu$ is a $\u1$ connection on the circle
with 2 singularities corresponding to the holonomy. Note that here $\mu_2=-
\mu_1=\omega$, with $\omega\equiv|\vec\omega|\in[0,\half]$. The magnetic 
components of $\hat{F}$ vanish and hence non-zero values and singularities are 
only assumed by the electric components $\hat{F}_{0i}$. By the Nahm equations
$\hat A$ is forced to be piecewise constant. We will give the explicit ADHM 
construction for these calorons and show among other things that all aspects 
suggested in ref.~\cite{NahMonCal} arise automatically.

\section{The ADHM construction}
We first summarise the ADHM construction~\cite{AtiDriHitMan,AtiFerLec} for 
charge $k$ $\su2$ instantons on $\re^4$. Generalisation to higher groups is 
well-known but will detract one from the simplicity of our construction. The 
ADHM data consist of a quaternionic row vector $\lambda=(\lambda_1,\cdots,
\lambda_k)$ and a quaternionic, symmetric $k \times k$ matrix $B$ ($\lambda_m
\equiv\lambda^\mu_m\sigma_\mu$ and $B_{m,n}\equiv B^\mu_{m,n}\sigma_\mu$, 
with $\lambda_m^\mu\in\re$ and $B_{m,n}^\mu=B_{n,m}^\mu\in\re$).
These objects are comprised in 
\beq
\Delta(x)=\left(\!\!\bea{c}\lambda\\B-x\eea\!\!\right).
\eeq
Here, the quaternion $x\equiv x_\mu\sigma_\mu$ denotes the position variable 
and a $k \times k$ unit matrix is implicit in our notation. For later use,
we define $\Delta\equiv\Delta(x=0)$. The ADHM gauge potential is given by
\beq
A_\mu(x)=v^\dagger(x)\partial_\mu v(x),\label{eq:AADHM}
\eeq
where $v(x)$ is a $(k+1)$ dimensional quaternionic vector, the normalised 
solution to
\beq
\Delta^\dagger(x)v(x)=0.\label{eq:deltav}
\eeq
For this construction to give a \sd\ potential $A_\mu$, $\Delta(x)$ has to
satisfy the ADHM constraints. These demand that $\Delta^\dagger(x)\Delta(x)$
be real quaternionic ({\ie}\ commutes with the quaternions) and invertible.
It is sufficient for this to hold at $x = 0$, \ie, 
$\Delta^\dagger\Delta=B^\dagger B+\lambda^\dagger\lambda$ must be real
quaternionic and invertible. Furthermore $(B-x)$ should have a trivial 
kernel, except for $k$ values of $x$, where $v(x)$ and, as a consequence, 
$A_\mu(x)$ are singular. This can be shown to be a gauge singularity and 
implements the non-triviality of the bundle and reflects the topological 
nature of these solutions.

This construction gives all instantons on $\re^4$. The following transformations
\beqa
&\lambda\rightarrow\lambda T^{-1},~B\rightarrow TBT^{-1},~T\in O(k),
\label{eq:adhmsym}&\\&\lambda\rightarrow g\lambda, \quad g\in\su2,\nonumber&
\eeqa
both leave the quadratic ADHM constraint untouched. The first does not
change $A_\mu(x)$, whereas the second induces a global gauge transformation. 
Local gauge transformations arise from the $U(2)$ symmetry $v(x)\rightarrow 
v(x)g(x)$ in the solution space of \refeq{deltav}. One must divide out these 
symmetries in order to obtain all gauge inequivalent solutions. This reduces 
the dimension of the space of gauge inequivalent solutions to $8k(-3)$, 
depending on whether or not the global gauge degrees of freedom are included as
moduli. Considering the $g$ as moduli, the moduli space is an $8k$ dimensional
hyperK\"ahler manifold~\cite{DonKro}.

Many aspects featuring in the construction above have their counterparts in the
Nahm transformation. The reality constraint is similar to the vanishing of the
imaginary quaternions in the Weitzenb\"ock formula, which leads to the \sd ity
of the Nahm connection. The symmetries in the ADHM construction can be traced
back to the triviality of the gauge action in the Nahm transformation and the
unitary symmetry of the fermionic zero modes. We define two matrix inverses, 
the analogues of the Green's functions,
\beq
f_x=(\Delta(x)^\dagger \Delta(x))^{-1}\in\re^{k \times k},\quad G_x=(
(B-x)^\dagger(B-x))^{-1}\in\qu^{k\times k},\label{eq:deffg}
\eeq
and a scalar function
\beq
\phi(x)=1+\lambda G_x\lambda^\dagger.\label{eq:norm}
\eeq
These Green's functions $f_x$ and $G_x$ are related, as can be seen from
the expansion of $f_x$ in terms of $G_x$,
\beq
f_x=(G_x^{-1}+\lambda^\dagger \lambda)^{-1}=G_x-G_x \lambda^\dagger
\sum_{n=0}^\infty(-\lambda G_x\lambda^\dagger)^n\lambda G_x=G_x-\phi^{-1}(x)
G_x \lambda^\dagger\lambda G_x,\label{eq:relfandG}
\eeq
Acting on \refeq{relfandG} with $\lambda^\dagger$ on the right and/or with
$\lambda$ on the left, yields
\beq
G_x\lambda^\dagger=\phi(x) f_x\lambda^\dagger, \quad\phi(x)=(1-\lambda f_x
\lambda^\dagger)^{-1}. \label{eq:grfurel}
\eeq
To solve for $v(x)$ in \refeq{deltav}, we introduce (a matrix of spinors) 
$u(x)$, and obtain
\beq
v(x)=\phi^{-\half}(x)\left(\!\!\bea{c}-1\\u(x)\eea\!\!\right),\quad
u(x)=(B^\dagger-x^\dagger)^{-1}\lambda^\dagger,\label{eq:vu}
\eeq
where $\phi(x)=1+\lambda G_x\lambda^\dagger=1+u^\dagger(x) u(x)$
accounts for the normalisation of $v(x)$. In terms of these quantities, the
gauge potential reads
\beq
A_\mu(x)=\half\phi^{-1}(x)(u^\dagger(x)\partial_\mu u(x)-\partial_\mu
u^\dagger(x) u(x)).\label{eq:Au}
\eeq
To show that the connection is indeed \sd\ is best seen from using 
$F=dA+A\wedge A$, where $A=A_\mu dx_\mu$ and $F=\half F_{\mu\nu}dx_\mu\wedge
dx_\nu$. With $A=v^\dagger(x) dv(x)$ one finds 
\beq
F=dv^\dagger(x)\wedge dv(x)-dv^\dagger(x) v(x)\wedge v^\dagger(x) dv(x)=
dv^\dagger(x)(1-v(x)\!\otimes\!v^\dagger(x))dv(x).
\eeq
As $1-v(x)\!\otimes\!v^\dagger(x)$ is the projection on the orthogonal 
complement of the kernel of $\Delta^\dagger(x)$ (since $\Delta^\dagger(x)
v(x)=0$), we can use that $1-v(x)\otimes v^\dagger(x)=\Delta(x)f_x
\Delta^\dagger(x)$. Substituting this in the expression for $F$ and using 
that $\Delta^\dagger(x)dv(x)=-dx_\mu\frac{\partial\Delta^\dagger(x)}{\partial 
x_\mu}v(x)=dx^\dagger(b^\dagger v(x))$ ($dx\equiv\sigma_\mu dx_\mu$,
$b^\dagger v(x)\equiv \phi^{-\half}(x)u(x)$), we find~\cite{AtiFerLec}
\beq
F=(v^\dagger(x) b)dx\wedge f_x dx^\dagger(b^\dagger v(x)).
\eeq
The crucial observation is now that the quadratic ADHM constraint implies that
$f_x$ commutes with the quaternions and that $dx\wedge dx^\dagger=\eta_{\mu\nu}
dx_\mu\wedge dx_\nu$, much like in the Nahm transformation where the 
Weitzenb\"ock identity, \refeq{Weitzenboeck}, guarantees that $D_z^\dagger(A)
D_z(A)$ commutes with the quaternions. We thus find~\cite{CorGod}
\beq
F_{\mu\nu}(x)=2\phi^{-1}(x)u^\dagger(x)\eta_{\mu\nu}f_xu(x),\label{eq:fdual}
\eeq
which is \sd\ due to the \sd ity of $\eta_{\mu\nu}=\sigma_{[\mu}\bsigma_{\nu]}$.

In the case at hand, \refeq{Au} is of little practical use as it stands. We
therefore rearrange it such that we can express $A_\mu$ in terms of evaluations
of the Green's function $f_x$. Using $\partial_\mu(B^\dagger-x^\dagger)^{-1}=
(B^\dagger-x^\dagger)^{-1}\bsigma_\mu(B^\dagger-x^\dagger)^{-1}=
(B-x)G_x\bsigma_\mu(B-x)G_x$, we get
\beq
A_\mu(x)=\phi^{-1}(x)~\lambda G_x\etabar_{\mu\nu}(B-x)_\nu G_x \lambda^\dagger.
\eeq
We substitute $G_x\lambda^\dagger=\phi(x)f_x\lambda^\dagger$, \refeq{grfurel},
and noting that
\beq
\partial_\nu f_x^{-1}=\partial_\nu G_x^{-1}=-2(B-x)_\nu,\quad
(\partial_\mu f_x^{-1})f_x=-f_x^{-1}\partial_\mu f_x,\label{eq:dfandG}
\eeq
we arrive at the following compact result for the gauge potential (see
also ref.~\cite{Temp})
\beq
A_\mu(x)=\half\phi(x)\partial_\nu\left(\lambda\etabar_{\mu\nu}f_x
\lambda^\dagger\right),
\eeq
using once again that $f_x$ commutes with the quaternions. When
$\etabar_{\mu\nu}$ is moved through $\lambda$, one finds an expression 
for $A_\mu$ in terms of (derivatives of) ``expectation values'' of the
Green's function $f_x$.
\beq
A_\mu(x)=\half\phi(x)\sigma_\alpha\etabar_{\mu\nu}\bar\sigma_\beta
\partial_\nu\phi_{\alpha\beta},\label{eq:gentH}
\eeq
where
\beq
\phi_{\alpha\beta}(x)=\phi_{\beta\alpha}(x)=(\lambda_\alpha f_x\lambda_\beta^t).
\label{eq:defscal}
\eeq
At this point we can make contact with the well-known 't Hooft 
ansatz~\cite{JacNohReb}. This forms a subclass of the ADHM construction with 
$\lambda$ real ($\lambda_m=\sigma_0\rho_m$) and $B_{m,n}=\delta_{m,n}y_m$ 
diagonal, corresponding to $k$ instantons with scales $\rho_m$ at positions 
$y_m$. This $5k$ dimensional family trivially satisfies the ADHM constraints. 
In this simpler situation the gauge potential can be written even in terms of 
a {\em single} scalar potential $\phi(x)=1+\sum_k\rho_k^2/|x-y_k|^2$ as 
$A_\mu(x)=\half\etabar_{\mu\nu}\partial_\nu\log\phi(x)$, since $\phi_{00}(x)=
1-\phi^{-1}(x)$. 

For the action density, the following expression can be found in the 
literature~\cite{Osb,CorGod}
\beq
\tr F^2_{\mu\nu}(x)=-\partial^2_\mu\partial^2_\nu\log\det f_x\label{eq:acdens}.
\eeq
This expression is regular everywhere. We can rewrite \refeq{acdens} using
\refeq{relfandG} and \refeq{dfandG} as
\beq
\tr F^2_{\mu\nu}(x)=-\half\partial^2_\mu\partial^2_\nu\log\det G_x+
\partial^2_\mu\partial^2_\nu\log\phi(x).\label{eq:acdensth}
\eeq
The factor $\half$ is due to $G_x$ being considered as a quaternionic and $f_x$
as a real $k\times k$ matrix. For the 't Hooft ansatz, $\partial^2_\mu\log
\det G_x$ vanishes, except for delta functions at $x=y_k$, and we retrieve 
the known result~\cite{JacNohReb}, $\tr F^2_{\mu\nu}=\partial^2_\mu
\partial^2_\nu\log\phi(x)$, which is indeed singular at these points 
(inadvertently in sect.~2 of ref.~\cite{PLB}, $\tr F_{\mu\nu}^2$ was 
given with the wrong sign).

\section{The construction of the caloron}
In this section we describe the ADHM construction of caloron solutions with
non-trivial holonomy. This will be a two-step process. Crucial will be the 
interpretation of the ADHM data as the Fourier coefficients of the Weyl 
operator in the Nahm transformation.  In our strategy, we build the caloron 
as an infinite, periodic (gauge-twisted) chain of instantons. It will be 
shown how we can realise this within the ADHM construction, by solving the 
quadratic constraint on the ADHM data. To find $A_\mu(x)$ we use again a 
Fourier transform to construct $\hat f_x(z,z')$, the Green's function of an 
ordinary second order differential equation, which allows for the determination 
of $\phi_{\alpha\beta}(x)=(\lambda_\alpha f_x \lambda_\beta^t)$, see 
\refeq{gentH}.

The boundary conditions $A_\mu(x+\cT)=e^{2\pi i\vec\omega\cdot\vec\tau}A_\mu(x)
e^{-2\pi i\vec\omega\cdot\vec\tau}$ are satisfied when
\beq
u_k(x+\cT)=u_{k-1}(x)\exp(-2\pi i\vec\omega\cdot\vec\tau),
\eeq
as is seen from \refeq{Au}. This is implemented by the periodicity constraints
\beq
\lambda_{k+1}=e^{2\pi i\vec\omega\cdot\vec\tau}\lambda_k,\quad
B(x+\cT)_{m,n}=B(x)_{m-1,n-1},\label{eq:percon}
\eeq
where $B(x)=B-x$. It now follows that
\beq
B_{m+1,n+1}=B_{m,n}+\cT\delta_{m,n},
\eeq
the inhomogeneous part of which is solved by having $\ldots,-2\cT,-\cT,0,\cT,
2\cT,\ldots$ on the diagonal of $B$. We still have to determine the remainder
of $B$, called $\hat A$ (anticipating its interpretation as Nahm connection), 
that contains its off-diagonal entries. In order to satisfy \refeq{percon}, 
$\hat A$ has to be of a convolutive type $\hat A_{m,n}=\hat A_{m-n}$, such that
\beq
\lambda_k=e^{2\pi ik\vec\omega\cdot\vec\tau}\zeta,\quad\zeta=\rho q,\quad
B_{m,n}=\cT m\delta_{m,n}+\hat A_{m-n}.\label{eq:param}
\eeq
Here $\zeta$ is an arbitrary quaternion. Its length $\rho=|\zeta|$ is the scale
parameter of the caloron. The $\su2$ element $q=\zeta/\rho$ describes its 
combined spatial and gauge orientation. The diagonal of $\hat A_{m,n}$ is 
necessarily constant, $\hat A^{\rm diag}_{m,n}\equiv\xi$, and plays 
the role of the position of the caloron. The ADHM data can now be readily 
interpreted as describing a periodic array of instantons, with temporal spacing
$\cT$ and relative gauge orientation $e^{2\pi i\vec\omega\cdot\vec\tau}$, with 
off-diagonal terms to account for the non-linear constraints. To simplify 
notations, we use the scale invariance of the self-duality equations to set 
$\cT=1$.  On dimensional grounds one can easily reinstate the proper $\cT$ 
dependence when required.

When we perform the Fourier transformation, $B$ will be transformed into a Weyl
operator, $\lambda$ and $\lambda^\dagger\lambda$ into delta function
singularities and $u(x)$ into a spinor (to be more precise a $2\times 2$ matrix
with as columns $\hat\psi_x^i$, cmp. sect.~2):
\beqa
&&\sum_{m,n}B_{m,n}(x)e^{2\pi i(mz-nz')}=\frac{\delta(z-z')}{2\pi i}\hat D_x
(z'),\quad\hat D_x(z)=\sigma_\mu D_x^\mu(z)=\ddz+\hat A(z)-2\pi ix,\nonumber\\
&&\sum_m\lambda_m e^{2\pi imz}=\hat\lambda(z),\quad\hat\lambda(z)=(P_+\delta(z-
\omega)+P_-\delta(z+\omega))\zeta,\quad P_\pm=\half(1\pm\hat\omega\cdot\vec
\tau),\nonumber\\&&\sum_m\lambda^\dagger_m\lambda_n e^{2\pi i(mz-nz')}=
\hat\lambda^\dagger(z')\hat\lambda(z)=\delta(z-z')\hat\Lambda(z),\quad
\hat\Lambda(z)=\bar\zeta\hat\lambda(z)=\hat\lambda^\dagger(z)\zeta,\nonumber\\&&
\sum_mu_m(x)e^{2\pi imz}=\hat\psi_x(z).\label{eq:fourtr}
\eeqa
All these objects are defined on $S^1$, or more appropriately from the Nahm 
perspective, on $\re^4/\hat H=\re^4/(\re^3\times\zahlen)$. Note that 
$\hat A(z)=\sigma_\mu\hat A_\mu(z)=2\pi i\sum_m\exp(2\pi imz)\hat A_m$, such 
that from the symmetry of $\hat A_{m,n}$ (implying $\hat A_m=\hat A_{-m}$) it
follows that $\hat A_\mu(z)$ is imaginary such that the differential operator 
$\hat D_x(x)$ is exactly the dual Weyl operator $\hat D_x(\hat A)$ introduced 
in sect. 2 (to agree with the notation there we have - unlike in ref.~\cite{PLB}
- included a factor $2\pi i$ in our definitions). Combining these features, we 
can interpret the Fourier transform of the ADHM construction as the inverse 
(or second) Nahm transformation.

The symmetries in the ADHM construction for periodic instantons lead to a $U(1)$
gauge symmetry for $\hat A_\mu(z)$. In order for \refeq{adhmsym} to preserve 
the periodicity constraint \refeq{percon}, $T$ has to be of a convolutive type,
$T_{m,n}=T_{m-n}$. Defining the periodic function $\hat g(z)=\sum_m\exp(2\pi 
imz)T_m$ and using the fact that $T$ is orthogonal ($T^{-1}_{m,n}=T_{n-m}$ and 
$\sum_k T_{k+n}T_k=\delta_{n,0}$), one concludes that $\hat g(z)\in U(1)$. A 
gauge transformation with $\hat g(z)$ leaves $\hat A_i(z)$ invariant and 
transforms $\hat A_0(z)$ to $\hat A_0(z)-d\log \hat g(z)/dz$. Note that 
$\hat A_0(z)$ can be gauged away, apart from a constant (as the Polyakov loop 
is gauge invariant). Hence we may choose $\hat A_0(z)=2\pi i\xi_0$. 

To implement the quadratic ADHM constraint after Fourier transformation we note
that any complex $2\times 2$ matrix $W$ can be decomposed as $W=W_\mu\sigma_\mu$
and that $[W,\sigma_\mu]=0$ can be implemented by requiring $\myIm W=0$, 
provided we define $\myIm W\equiv\half[W-\tau_2W^t\tau_2]$. Also note that 
$\myRe W\equiv\half[W+\tau_2W^t\tau_2]=\half\tr~W=W_0$. Thus, the quadratic 
ADHM constraint can be formulated as
\beq
\myIm(\hat D_x^\dagger(z)\hat D_x(z)+4\pi^2\hat\Lambda(z))=0.\label{eq:ftadhmc}
\eeq
From the Weitzenb\"ock formula $\hat D^\dagger_x(z)\hat D_x(z)=-(
\hat D_x^\mu(z)\hat D_x^\mu(z)+\half\etabar_{\mu\nu}\hat F_{\mu\nu})$ 
(cmp.~\refeq{Weitzenboeck}) we find
\beqa
\ddz\hat A(z)&=&\half\etabar_{\mu\nu}\hat F_{\mu\nu}(z)=-\myIm\hat D^\dagger_x
(z)\hat D_x(z)=4\pi^2\myIm\hat\Lambda(z)\nonumber\\&=&2\pi^2(\bar\zeta\hat
\omega\cdot\vec\tau\zeta)(\delta(z-\omega)-\delta(z+\omega)),\label{eq:curvsing}
\eeqa
using \refeq{ftadhmc} and $\hat F_{ij}(z)=0$. This leads to~\cite{Kraan}
\beq
\hat A(z)=\sigma_\mu\hat A_\mu(z)=2\pi i[\xi+\pi(\bar\zeta\hat\omega\cdot
\vec\sigma\zeta)\Theta_\omega(z)],\label{eq:nahmdata}
\eeq
where (see fig.~1)
\beq
\Theta_\omega(z)=(\chi_{[-\omega,\omega]}(z)-2\omega).\label{eq:theta}
\eeq
Here $\chi_{[a,b]}(z)=1$ if $z\in[a,b]$ and 0 elsewhere, properly defined on
the circle. We have arranged $2\pi i\xi=\int dz~\hat A(z)$, 
such that $B(x)$ has a single zero mode for $x=\xi$, to agree with the 
interpretation of $\xi$ as the position (centre of mass) of the caloron. 
Fourier transforming back, we retrieve the matrix representation of $B(x)$
\beq
B_{m,n}(x)=(m+\xi-x)\delta_{m,n}+\bar\zeta\hat{\omega}\cdot\vec\sigma\zeta
\frac{\sin(2\pi(m-n)\omega)}{m-n}(1-\delta_{m,n}).\label{eq:ADHMdata}
\eeq
The moduli space is thus parametrised by the caloron position $\xi$ 
and by its scale and orientation $\zeta=\rho q$, with $\xi_0\in S^1$, 
$\vec\xi\in\re^3$, $\rho\in\re^+$ and $q\in\su2$.

\begin{figure}[htb]
\vspace{3.6cm}
\includegraphics{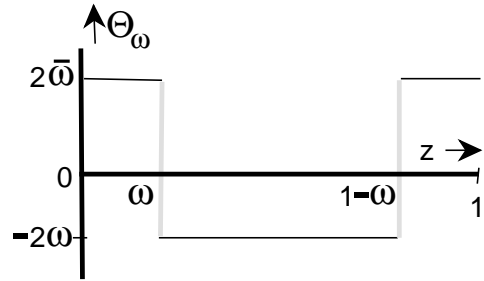}
\caption{The function $\Theta_\omega(z)$.}
\end{figure}

We end this first step in the construction by noting that the delta function 
singularities arise precisely as predicted by the general properties of the 
Nahm transformation, discussed in section 2 and that $\hat A(z)$ constructed 
in this section solves the Nahm equations, whereby we have found all \sd\ 
solutions for $\re^3\times S^1$.

For the second step we have to find the Green's function $f_x$. For further 
notational simplification we absorb $\xi$ by a translation (we have already 
used the scale invariance to fix $\cT=1$) such that after a \FT\ the definition 
of $f_x$, \refeq{deffg}, can be cast into a differential equation for 
$\hat f_x(z,z')\equiv\sum_{n,m}f_x^{m,n}e^{2\pi i(mz-nz')}$
\beq
\left\{\!\!\left(\!\frac{1}{2\pi i}\ddz\!-\!x_0\right)^{\!\!2}\!\!+\!s^2
\chi_{[-\omega,\omega]}(z)\!+\!r^2\chi_{[\omega,1-\omega]}(z)\!+\!
\frac{\rho^2}{2}(\delta(z\!\!-\!\!\omega)\!+\!\delta(z\!\!+\!\!
\omega))\!\right\}\!\!\hat f_x(z,z')=\delta(z-z').\label{eq:wa}
\eeq
Here, the radii $r$ and $s$ are given by
\beq
r^2=\half\tr(\vec x\cdot\vec\tau+2\pi\omega\rho^2\bar q\hat\omega\cdot\vec\tau 
q)^2,\quad s^2=
\half\tr(\vec x\cdot\vec\tau-2\pi\bar\omega\rho^2\bar q\hat\omega\cdot\vec\tau 
q)^2,\quad(\bar\omega=\half-\omega),\label{eq:radii}
\eeq
and can be interpreted as the respective centre of mass radii of the two
constituent monopoles of the caloron. Note that $\bar q\hat\omega\cdot\vec\tau
q$ shows how global gauge rotations are to be correlated to spatial rotations 
so as to keep the holonomy unchanged. We will come back to this in the next 
section. The symmetries of $\Delta^\dagger(x)\Delta(x)$ imply for the Green's 
function $\hat f_x(z,z')$ the following relations,
\beq
\hat f_x(z,z')=\hat f_x(-z, -z')^*=\hat f_x(-z',-z)=\hat f_x(z',z)^*.
\eeq
In particular we have $\hat f_x(\omega,\omega)=\hat f_x(-\omega,-\omega)\in\re$
and $\hat f_x(\omega,-\omega)=\hat f_x(-\omega,\omega)^*$.
The Green's function is that of ordinary quantum mechanics on a circle
with a piecewise constant potential and delta function singularities of 
strength $\half\rho^2$ at the jumping points $z=\pm\omega$. It can thus be
constructed in the usual straightforward (but tedious) method of matching 
the value of the Green's function and its derivative (up to the appropriate
jumps) at $z=\pm\omega$ and $z=z'$. The result reads
\beqa
&&\hat f_x(z,z')=\chi_{[-\omega,\omega]}(z')\left(\chi_{[-\omega,\omega]}(z)\hat
f^d_x(z,z',r,s,\omega)+\chi_{[\omega,1-\omega]}(z)\hat f^o_x(z,z',r,s,\omega)
\right)+\nonumber\\&&\quad~\,\chi_{[\omega,1-\omega]}(z')\left(\chi_{[\omega,
1-\omega]}(z)\hat f^d_x(z\!-\!\half,z'\!-\!\half,s,r,\bar\omega)+\chi_{[-\omega,
\omega]}(z)\hat f^o_x(z',z,r,s,\omega)^*\right).
\eeqa
In the following the diagonal component $\hat f^d_x(z,z')$ is only defined 
strictly for $z,z'\in[-\omega,\omega]$ and the off-diagonal component 
$\hat f^o_x(z,z')$ only for $z\in[\omega,1-\omega]$ and $z'\in[-\omega,\omega]$.
For $z$ or $z'$ outside of these intervals, one first has to map back to the 
interval $[-\omega,1-\omega]$, using periodicity.
\beqa
&&\hskip-7mm\hat f^d_x(z,z',r,s,\omega)=e^{2\pi ix_0(z-z')}\pi(rs\psi)^{-1}
\Big\{e^{-2\pi ix_0{\rm sign}(z-z')}r\sinh(2\pi s|z\!-\!z'|)\nonumber\\&&\qquad
\qquad~~-s^{-1}\cosh(2\pi s(z\!+\!z'))\left[\pi\rho^2r\cosh(4\pi r\bar\omega)+
\half(r^2\!-\!s^2\!+\!\pi^2\rho^4)\sinh(4\pi r\bar\omega)\right]\nonumber\\&&
\qquad+s^{-1}\cosh(2\pi s(2\omega\!-\!|z\!-\!z'|))\left[\pi\rho^2r\cosh(4\pi r
\bar\omega)+\half(r^2\!+\!s^2\!+\!\pi^2\rho^4)\sinh(4\pi r\bar\omega)\right]
\nonumber\\&&\hskip4.3cm+\sinh(2\pi s(2\omega\!-\!|z\!-\!z'|))\left[r\cosh(4\pi 
r\bar\omega)+\pi\rho^2\sinh(4\pi r\bar\omega)\right]\Big\},\nonumber\\
&&\hskip-7mm\hat f^o_x(z,z',r,s,\omega)=e^{2\pi ix_0(z-z')}\pi(rs\psi)^{-1}
\Big\{\pi\rho^2\sinh(2\pi r(1\!-\!z-\!\omega))\sinh(2\pi s(z'\!+\!\omega))
\\&&~\,+r\cosh(2\pi r(z\!-\!1\!+\!\omega))\sinh(2\pi s(z'\!+\!\omega))
-s\sinh(2\pi r(z\!-\!1+\!\omega))\cosh(2\pi s(z'\!+\!\omega))\nonumber\\&&
+e^{-2\pi ix_0}\Big[s\sinh(2\pi r(z\!-\!\omega))\cosh(2\pi s(z'\!-\!\omega))
-r\cosh(2\pi r(z\!-\!\omega))\sinh(2\pi s(z'\!-\!\omega))\nonumber\\&&\quad
\hskip7.4cm -\pi\rho^2\sinh(2\pi r(z\!-\!\omega))\sinh(2\pi s(z'\!-\!\omega))
\Big]\Big\},\nonumber
\eeqa
where we have introduced the scalar function
\beqa
&&\psi=-\cos(2\pi x_0)+\cosh(4\pi r\bar\omega)\cosh(4\pi s
\omega)+\frac{(r^2\!+\!s^2\!+\!\pi^2\rho^4)}{2rs}\sinh(4\pi r
\bar\omega)\sinh(4\pi s\omega)\nonumber\\&&\qquad\qquad+\pi\rho^2
\left(s^{-1}\sinh(4\pi s\omega)\cosh(4\pi r\bar\omega)+r^{-1}\sinh(4\pi
r\bar\omega)\cosh(4\pi s\omega)\right) .
\eeqa
In particular,
\beqa
&&\hat f_x(\omega,-\omega)=\pi(rs\psi)^{-1}e^{4\pi ix_0\omega}
\left\{e^{-2\pi ix_0}r\sinh(4\pi s\omega)+s\sinh(4\pi r\bar\omega)\right\},\\
&&\hat f_x(\omega,\omega)\!=\!\pi(rs\psi)^{-1}\Big\{s\sinh(4\pi r\bar\omega)
\cosh(4\pi s\omega)\!+\!r\sinh(4\pi s\omega)\cosh(4\pi r\bar\omega)
\nonumber\\&&\hskip8cm+\pi\rho^2\sinh(4\pi r\bar\omega)\sinh(4\pi s\omega)
\Big\}.\nonumber
\eeqa
and (see \refeq{grfurel})
\beq
\phi(x)=(1-\lambda f_x\lambda^\dagger)^{-1}=(1-\rho^2\hat f_x(\omega,
\omega))^{-1}\equiv\psi/\hat\psi,\label{eq:valnorm}
\eeq
where
\beq
\hat\psi=-\cos(2\pi x_0)+\cosh(4\pi r\bar\omega)\cosh(4\pi s\omega)+\frac{(r^2
\!+\!s^2\!-\!\pi^2\rho^4)}{2rs}\sinh(4\pi r\bar\omega)\sinh(4\pi s\omega).
\label{eq:psihat}
\eeq

We now use eqs.~(\ref{eq:gentH}-\ref{eq:defscal}) to determine $A_\mu$. The 
scalar functions $\phi_{\alpha\beta}$ are all defined in terms of 
$\hat f_x(\omega,\pm\omega)$. To get compact expressions, we introduce the 
complex function
\beq
\chi=\rho^2\hat f_x(\omega,-\omega)=e^{4\pi ix_0\omega}\frac{\pi\rho^2}{\psi}
\left\{e^{-2\pi ix_0}s^{-1}\sinh(4\pi s\omega)+r^{-1}\sinh(4\pi r\bar\omega)
\right\}.
\eeq
We choose $\vec\omega$ in the $z-$direction and $q=1$, which can always be
achieved by performing a suitable gauge and spatial rotation. We find for 
those functions $\phi_{\alpha\beta}$ that are non-zero
\beq
\phi_{00}=\half(1-\phi^{-1}+\myre\chi),\quad\phi_{33}=\half(1-\phi^{-1}-\myre 
\chi),\quad\phi_{03}=\phi_{30}=\half\myim\chi,
\eeq
such that with the use of \refeq{gentH} (for a matrix $W$, $\myre W\equiv
\half(W+W^\dagger)$)
\beq
A_\mu=\frac{i}{2}\bar\eta^3_{\mu\nu}\tau_3\partial_\nu\log\phi+\frac{i}{2}
\phi\myre\left((\bar\eta^1_{\mu\nu}-i\bar\eta^2_{\mu\nu})(\tau_1+i\tau_2)
\partial_\nu\chi\right).\label{eq:solution}
\eeq
For $\omega=0$ this reduces to the Harrington-Shepard solution of the 
caloron with trivial holonomy, since in that case $\chi=1-\phi^{-1}$.

The \sd ity of \refeq{solution} follows from \refeq{fdual}, but has also been 
checked numerically. In the asymptotic regime of large distances $|\vec x|$, 
\beq
\hat f_x(z,z')\approx\frac{\pi}{|\vec x|}e^{-2\pi{|\vec x|}|z-z'|+
2\pi ix_0(z-z')},\label{eq:fas}
\eeq
(with $|z-z'|$ the obvious distance function on the circle) from which it 
follows that $A_\mu$ tends to zero at spatial infinity. The holonomy at 
spatial infinity is then fully carried by the cocycle, and equals $e^{2\pi i
\vec\omega\cdot\vec\tau}$ as required. The non-trivial holonomy becomes even 
more transparant in the periodic gauge by performing a gauge transformation 
$g(x)=e^{-2\pi ix_0\vec\omega\cdot\vec\tau}$. This yields
\beq
A_\mu^{\rm per}=\frac{i}{2}\bar\eta^3_{\mu\nu}\tau_3\partial_\nu\log
\phi+\frac{i}{2}\phi\myre\left((\bar\eta^1_{\mu\nu}-i\bar\eta^2_{\mu\nu})
(\tau_1+i\tau_2)(\partial_\nu+4\pi i\omega\delta_{\nu,0})\tilde\chi\right)
+\delta_{\mu,0}2\pi i\omega\tau_3,\label{eq:solutionpg}
\eeq
where
\beq
\tilde\chi\equiv e^{-4\pi ix_0\omega}\chi=\frac{\pi\rho^2}{\psi}\left\{e^{-2\pi
ix_0}s^{-1}\sinh(4\pi s\omega)+r^{-1}\sinh(4\pi r\bar\omega)\right\}.
\eeq
In this gauge we immediately read off the constant background field at spatial
infinity, $A^{\rm per}_\mu=2\pi i\vec\omega\cdot\vec\tau\delta_{\mu,0}$, 
responsible for the holonomy in the periodic gauge. This concludes the 
construction of the caloron solution.

To use \refeq{acdens} for the action density we have to regularise the 
determinant, which for calorons diverges. However, $\partial_\mu\log\det f_x$ 
turns out to be finite. With the help of \refeq{dfandG} we find
\beq
\partial_\mu\log\det f_x=\partial_\mu\Tr\,\log f_x=\frac{1}{\pi i}\Tr\,
\hat D^\mu_xf_x=\frac{1}{\pi i}\int_{S^1}dz~\lim_{z'\rightarrow z}
\hat D^\mu_x(z)\hat f_x(z,z'),
\eeq
where $\Tr$ denotes the Hilbert space trace. We use point-splitting to define
\beq
\lim_{z'\rightarrow z}\ddz \hat f_x(z,z')\equiv\half\left(\lim_{\epsilon
\downarrow 0}\ddz \hat f_x(z+\epsilon,z')+\lim_{\epsilon\downarrow 0}\ddz 
\hat f_x(z-\epsilon,z')\right)\Bigg|_{z'=z},
\eeq
in accordance with the Fej\'er theorem for the convergence of Fourier
series, see e.g. ref.~\cite{WhiWat}. Careful inspection shows that
\beq
\Tr\,\hat D^\mu_xf_x=-\pi i\partial_\mu\log\psi,
\eeq
leading to the following miraculously simple result
\beq
-\tr F_{\mu\nu}^2(x)=\partial_\mu^2\partial_\nu^2\log\det f_x=
-\partial_\mu^2\partial_\nu^2\log\psi.\label{eq:acdensres}
\eeq
Note that $\psi$ is positive definite and smooth, despite its appearance. The
same applies for \refeq{acdensres}. The action density $-\half\tr F_{\mu\nu}^2
(x)$ takes its maximal value at $x_0=0$. We have verified \refeq{acdensres} 
numerically, using \refeq{solution}. Since the action density is a total 
derivative, one can express the total action in terms of a surface integral 
at spatial infinity. Using that $\partial_\mu^2\log\psi=4\pi/|\vec x|+
\cO(|\vec x|^{-4})$, one easily verifies that for the topological 
charge
\beq
k=-\frac{1}{8\pi^2}\int\tr F\wedge F=-\frac{1}{16\pi^2}\int d_4 x~
\tr F^2_{\mu\nu}(x)=1.\label{eq:pontr}
\eeq

In the appendix we give the expression for the Green's function $G_x$, from 
which it follows that
\beq
\partial_\mu^2\partial_\nu^2\log\det G_x=-\half\partial_\mu^2\partial_\nu^2
\log\hat\psi,\label{eq:detG}
\eeq
in accordance with \refeq{acdensth} and \refeq{valnorm}.

\section{Properties of the caloron solution}
We first discuss the issue of orientations in colour and real space. 
In general only the center $Z_2$ of the group of global gauge transformations
will leave the gauge potential invariant. The framing - embedding of the
solution in colour space - is in general not invariant under global gauge 
rotations. For non-trivial holonomy ($\omega\bar\omega\neq0$, or $\cP\neq\pm 1$
at infinity), also the holonomy is not invariant under such global gauge 
rotations, except for a $U(1)$ subgroup generated by $\hat\omega\cdot\vec\tau$,
which in the monopole terminology generates the unbroken gauge symmetry. For 
each choice of the holonomy - which can not change under continuous 
deformations - we have a separate caloron parameter space. We note that the 
spatial orientation is given by the preferred axis that appears in the formula 
for the action density and in the definition of the two radii $r$ and $s$, 
\refeq{radii}. The action density has an axial symmetry around $\hat a$ 
defined through 
\beq
\hat a\cdot\vec\tau=\bar q\hat\omega\cdot\vec\tau q=\bar\zeta\hat\omega\cdot\vec
\tau\zeta/\rho^2.
\eeq
When $q$ is part of the $U(1)$ subgroup generated by $\hat\omega\cdot\vec\tau$,
it does not affect the orientation of the solution, and indeed can be pulled
through in \refeq{param}, to be identified with the global gauge invariance
of \refeq{adhmsym} associated to this residual $U(1)$. The dimension of the 
moduli space of {\em gauge inequivalent} solutions at fixed holonomy, including
the position of the caloron described by $\xi_\mu$, is thus 7 for non-trivial 
and 5 for trivial holonomy. A global residual $U(1)$ gauge transformation 
(or a global $SU(2)$ gauge transformation in case of trivial holonomy), 
however, does change the framing of the solutions and the moduli space of 
framed calorons is 8 dimensional. Including these global gauge degrees of 
freedom will reveal the hyperK\"ahler structure of the moduli space, to be 
discussed in the next section.

Since a global gauge transformation leaves $\hat a$ invariant, we can best 
describe the parameters of the solutions for the choice where $\hat\omega=
\hat e_3$, i.e. $\hat\omega$ is pointing in the positive $x_3-$direction. Due 
to the residual gauge group, to any point on the two sphere defined by the 
symmetry axes of the caloron solution, a full $U(1)$ can be associated. 
This gives the Hopf fibration of $S^3=SU(2)$ over $S^2$, the fiber being 
$U(1)$. Using Euler angles we may choose the parametrisation 
\beq
q=e^{-i\Upsilon\frac{\tau_3}{2}}e^{i(\frac{\pi}{2}-\theta)\frac{\tau_2}{2}}
e^{-i\varphi\frac{\tau_1}{2}},\quad 0\leq\Upsilon\leq4\pi,\quad0\leq\varphi
\leq2\pi,\quad0\leq\theta\leq\pi,\label{eq:Euler}
\eeq
which leads, for $\hat\omega=\hat e_3$, to the axis of axial symmetry
$\hat a=(\cos\theta,\sin\theta\sin\varphi,\sin\theta\cos\varphi).$
The variable $\Upsilon$ describes the residual $U(1)$ gauge group generated 
by $\tau_3(=\hat\omega\cdot\vec\tau)$. With $q=q_\mu\sigma_\mu$ ($|q|=1$) we 
can introduce the Maurer-Cartan one-forms
\beq
{\Sigma}_i=2\eta^i_{\mu\nu}q_\mu dq_\nu,\quad d\Sigma_i=
\half\epsilon_{ijk}\Sigma_j\wedge\Sigma_k.\label{eq:mcartan}
\eeq
In terms of the Euler angles these read
\beqa
\bea{l}
\Sigma_1=\cos\Upsilon\sin\theta d\varphi-\sin\Upsilon d\theta,\\
\Sigma_2=\sin\Upsilon\sin\theta d\varphi+\cos\Upsilon d\theta,\\
\Sigma_3=d\Upsilon+\cos\theta d\varphi.\\
\eea
\eeqa
\vskip1.5cm\hskip1cm $\omega=0$\hskip6.5cm$\omega=0.25$
\vskip-1.5cm
\begin{figure}[htb]
\vspace{5.3cm}
\includegraphics{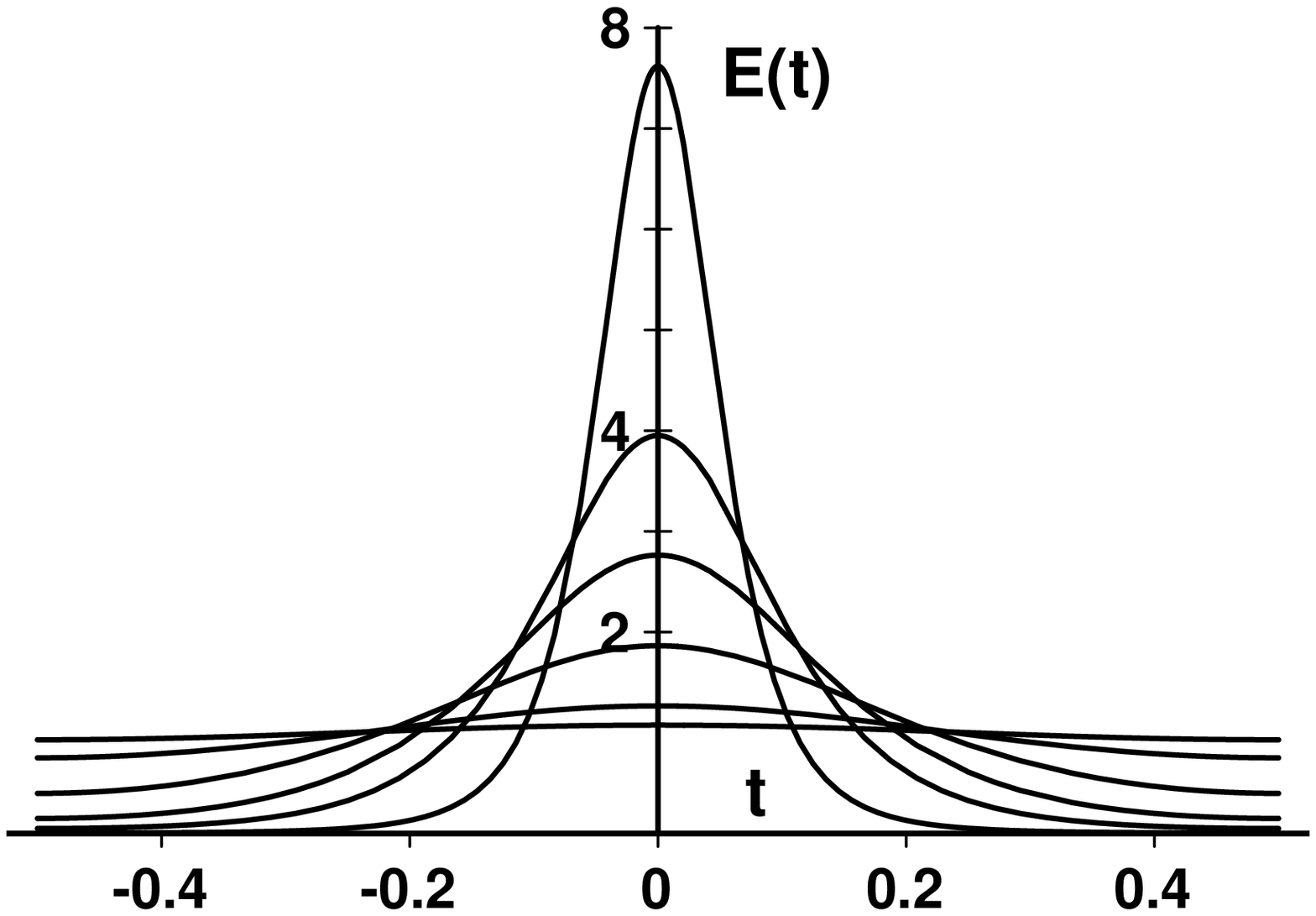}
\includegraphics{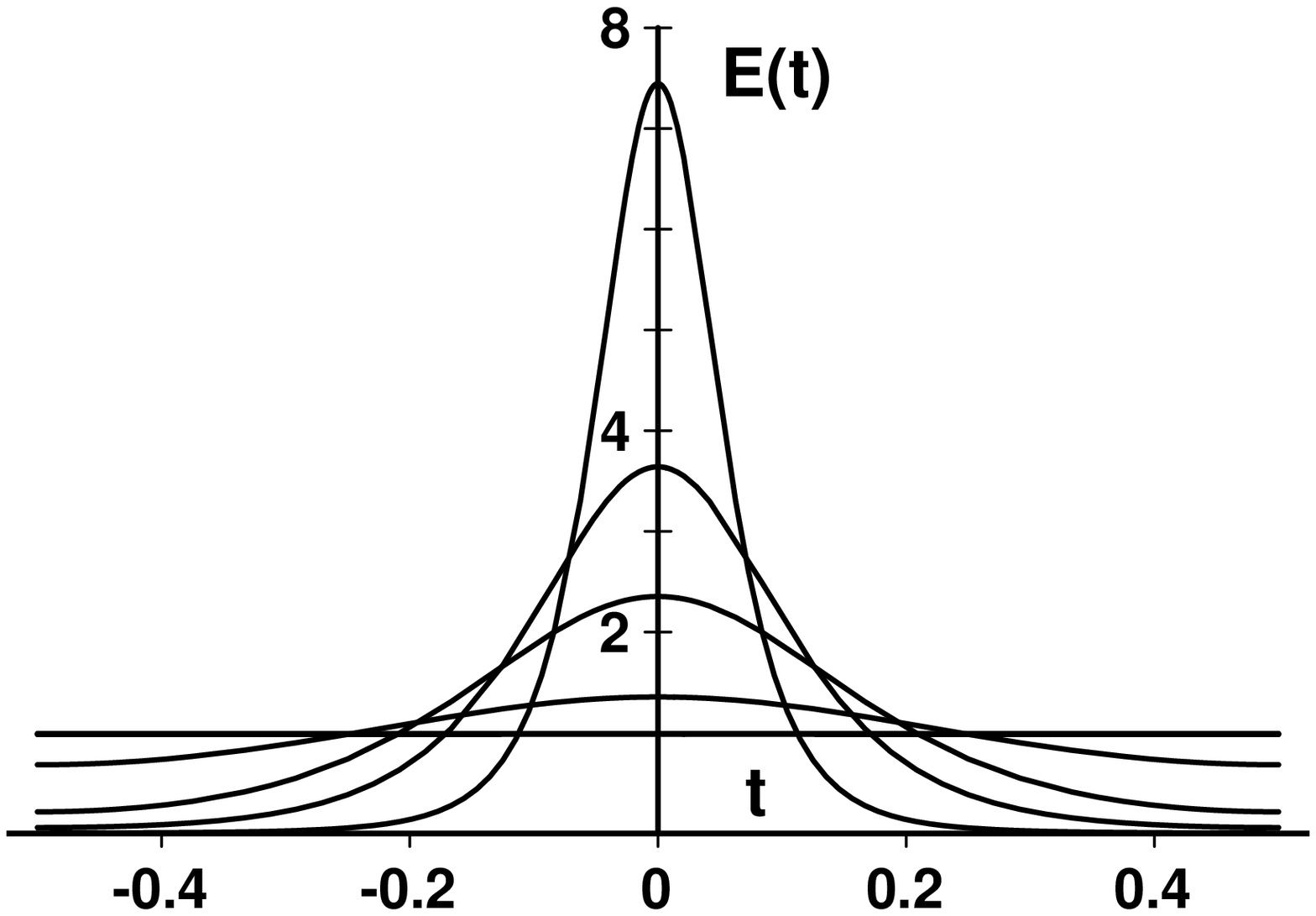}
\caption{Time evolution of the caloron solution.  During one period ($\cT=1$), 
we plot the ``energy" as a function of time, $E(t)\equiv-\frac{1}{16\pi^2}
\int_{\re^3}\tr F^2$, for $\rho=0.1,0.2,0.3,0.5,1.0,2.0$. For small values 
of $\rho$, the caloron is short-lived and instanton-like, whereas for large 
values, $\rho > 1$, the profile flattens and the caloron becomes static and 
monopole-like.}
\end{figure}

In order to visualise the caloron solution, we use \refeq{acdensres}. A 
time-slice of the caloron shows that it generically consists of two lumps. In 
fig.~2 we study the time dependence for 
various values of $\rho$. For small $\rho$ the caloron approaches the ordinary 
single instanton solution, with no dependence on $\omega$, as $\rho 
\rightarrow 0$ is equivalent to $\cT\rightarrow\infty$. Finite size effects 
set in when the size of the instanton becomes of the order of the 
compactification length $\cT$, i.e. when the caloron bites in its own tail. 
This occurs at roughly $\rho=\half\cT$. At this point, for $\omega\bar\omega
\neq0$ (i.e. the holonomy $\cP\neq\pm1$), two lumps are formed, whose 
separation grows as $\pi\rho^2/\cT$ (\cf\ \refeq{radii}). At large $\rho$ the 
solution spreads out over the entire circle in the euclidean time direction and
becomes static in the limit $\rho\rightarrow\infty$. So for large $\rho$ the 
lumps are well separated, see fig.~3. When far apart they become spherically
symmetric. As they are static and \sd\ they are necessarily BPS monopoles. 
\begin{figure}[htb]
\vspace{13.3cm}
\includegraphics{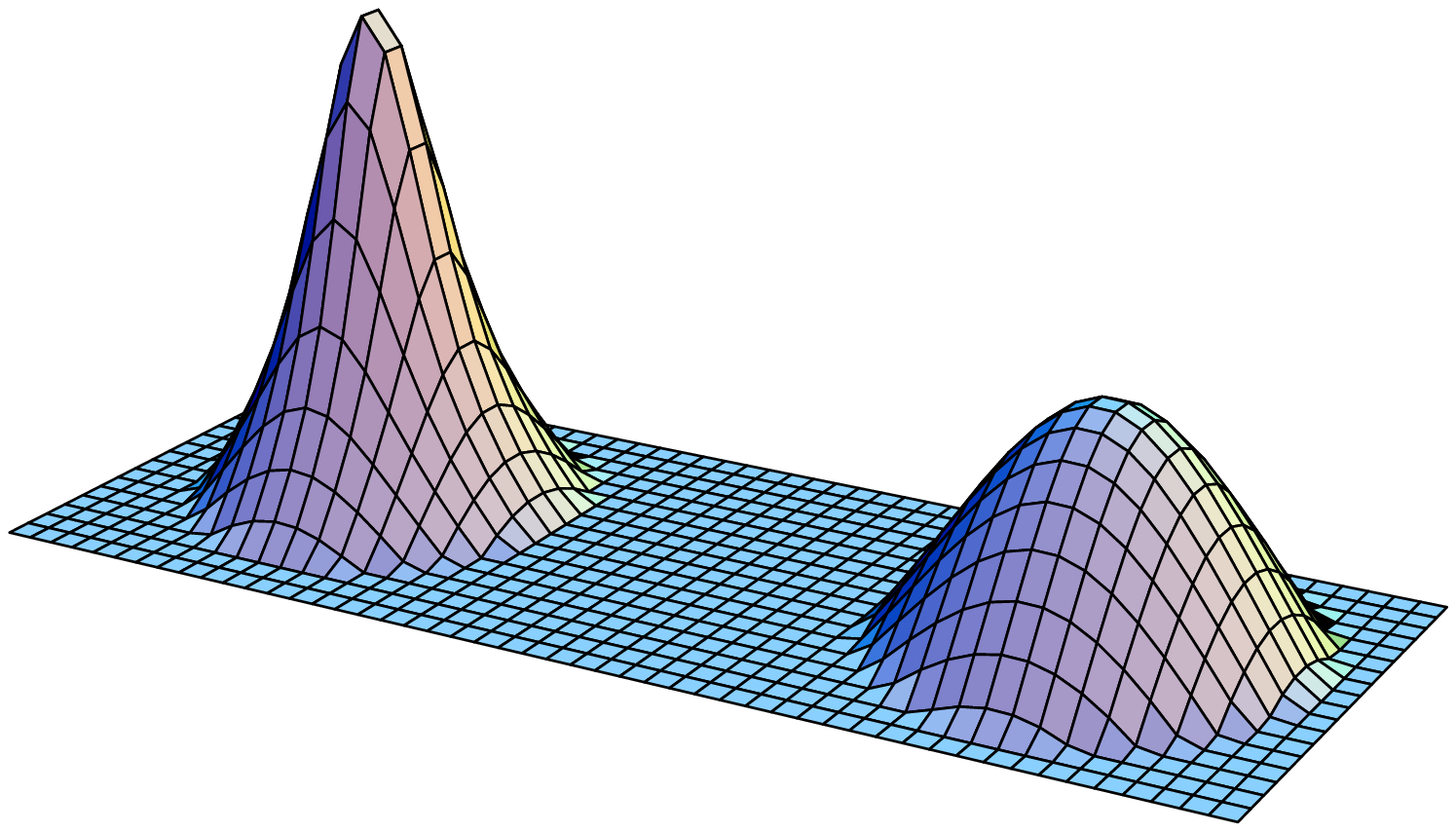}
\includegraphics{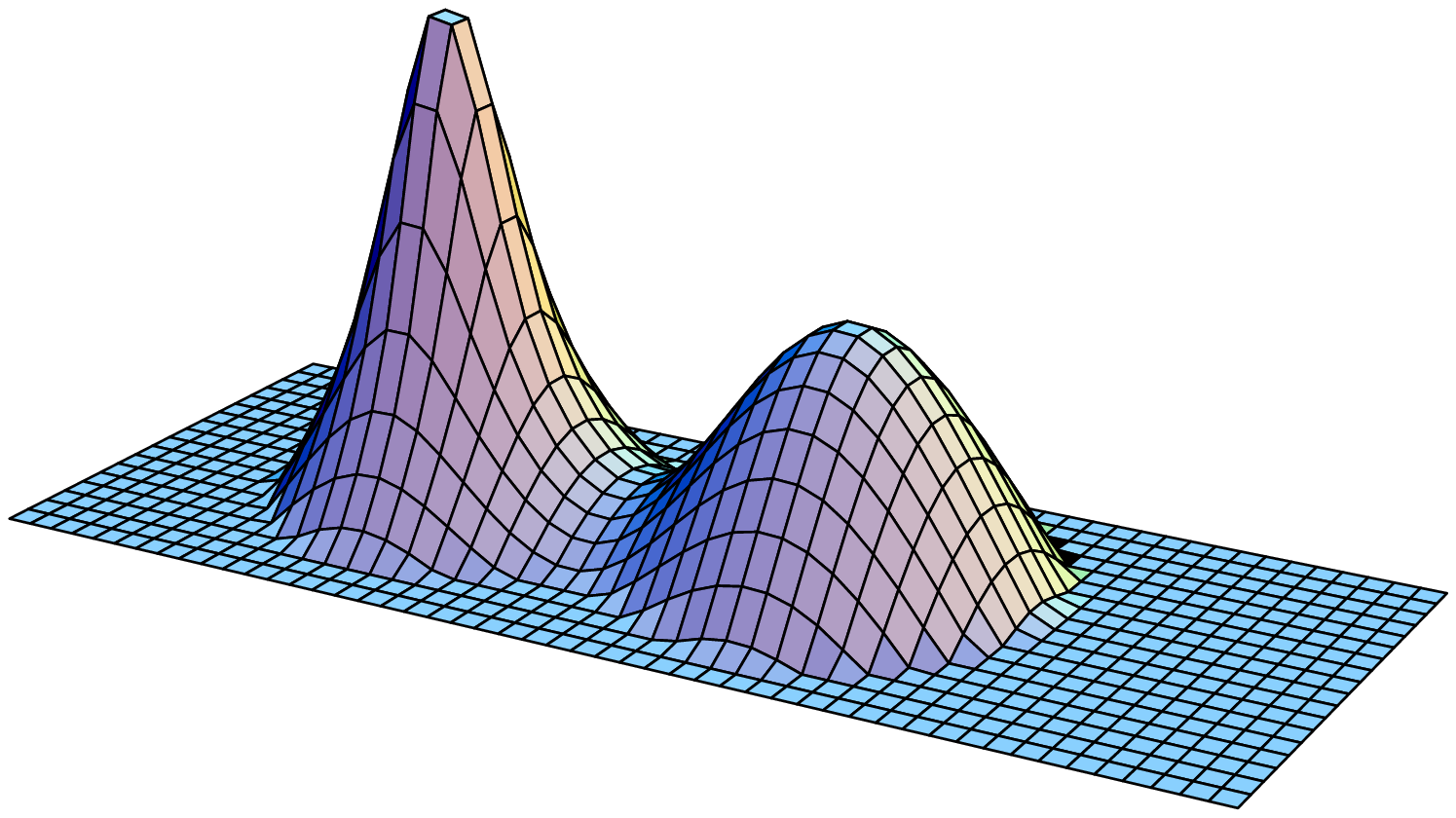}
\includegraphics{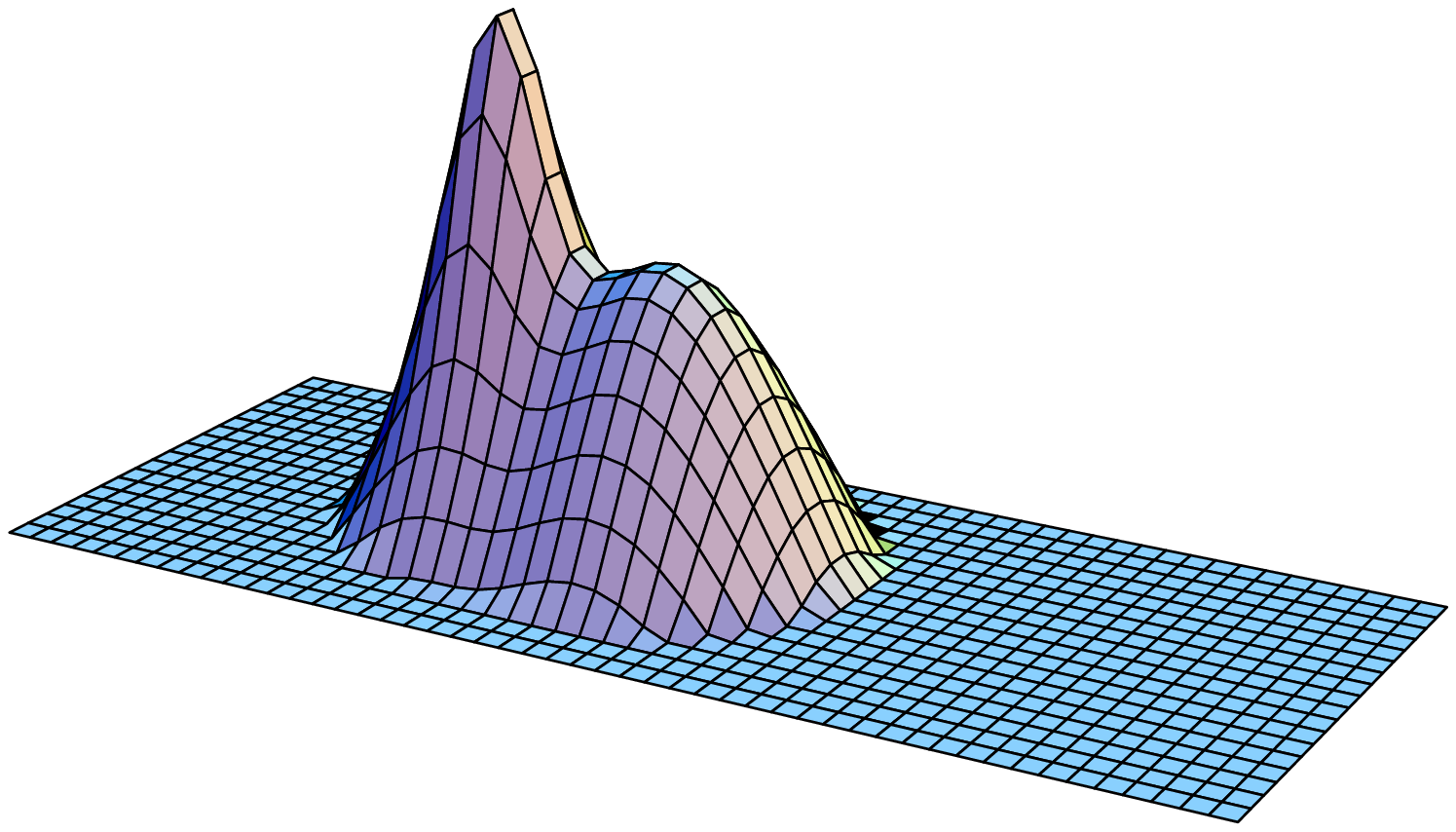}
\caption{Shown are caloron profiles for $\omega=0.125$ ($\cT=1$), with 
$\rho=0.8,1.2,1.6$ (from bottom to top). This illustrates the growing 
separation of the two lumps with $\rho$. Once the constituents are separated, 
the lumps are spherically symmetric and do not change their shape upon further 
separation. Vertically is plotted the action density at $x_0=0$, on equal 
logarithmic scales for all profiles. They were cut off at an action density 
below $1/e^2$.}
\end{figure}
One can show in this limit that they have unit, but opposite, magnetic 
charges and that the two lumps have spatial scales proportional to respectively 
$1/\bar\omega$ and $1/\omega$ (see sect.~7). Their action densities (or energy 
densities in this static limit) scale with $\bar\omega^4$ and $\omega^4$. After
integration, this results in monopole masses of respectively $16\pi^2\bar
\omega/\cT$ and $16\pi^2\omega/\cT$ for the two lumps, their mass ratio is 
therefore $\bar\omega/\omega$. The total energy, simply obtained by addition, 
indeed conforms with the unit topological charge of the solution. For 
$\omega=0$ or $\omega=\half$, the second lump is absent and the solution is 
spherically symmetric. For generic $\omega$, the solution has only an axial 
symmetry around the axis $\hat a$ connecting the two lumps. For 
$\omega=\quarter$, the lumps are equally sized, and the solution has a mirror 
symmetry in the plane perpendicular to $\hat a$, see fig.~4. 

\begin{figure}[htb]
\vspace{11cm}
\includegraphics{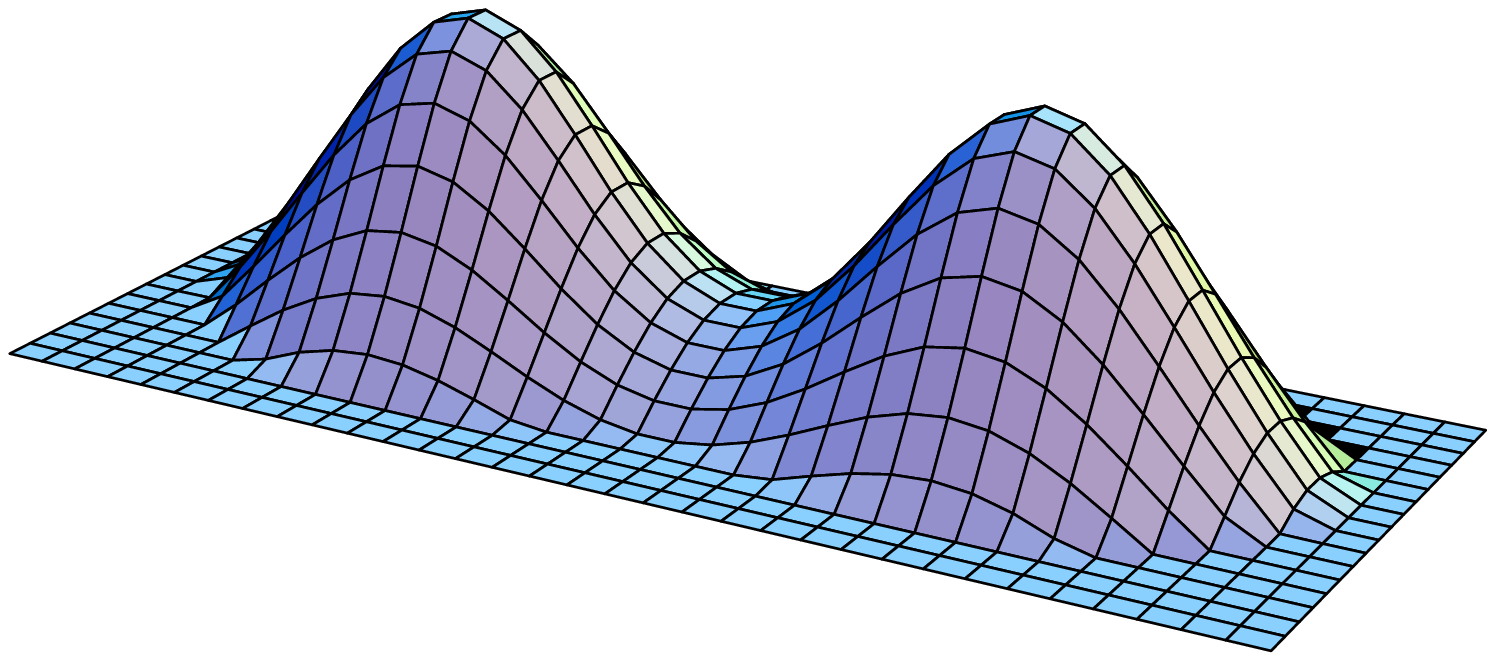}
\includegraphics{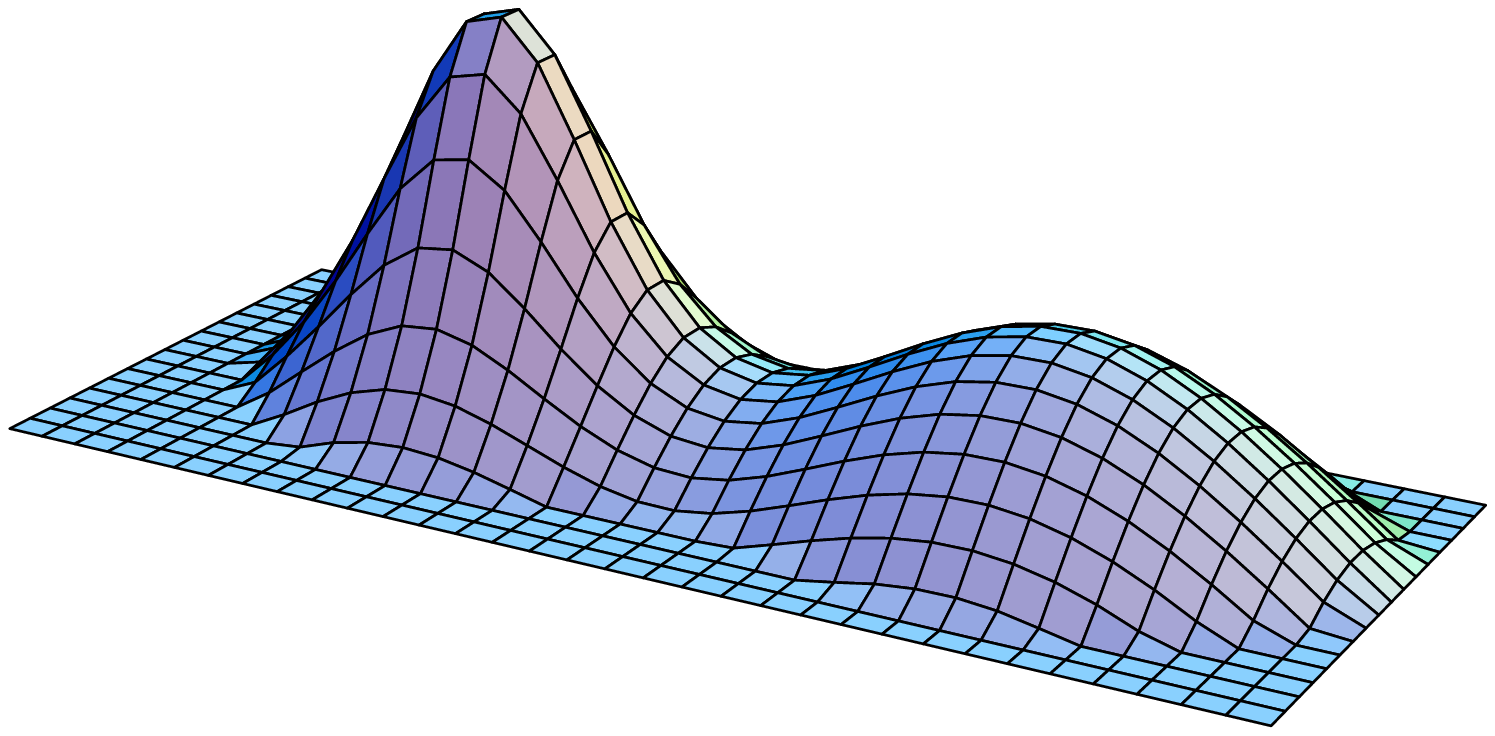}
\includegraphics{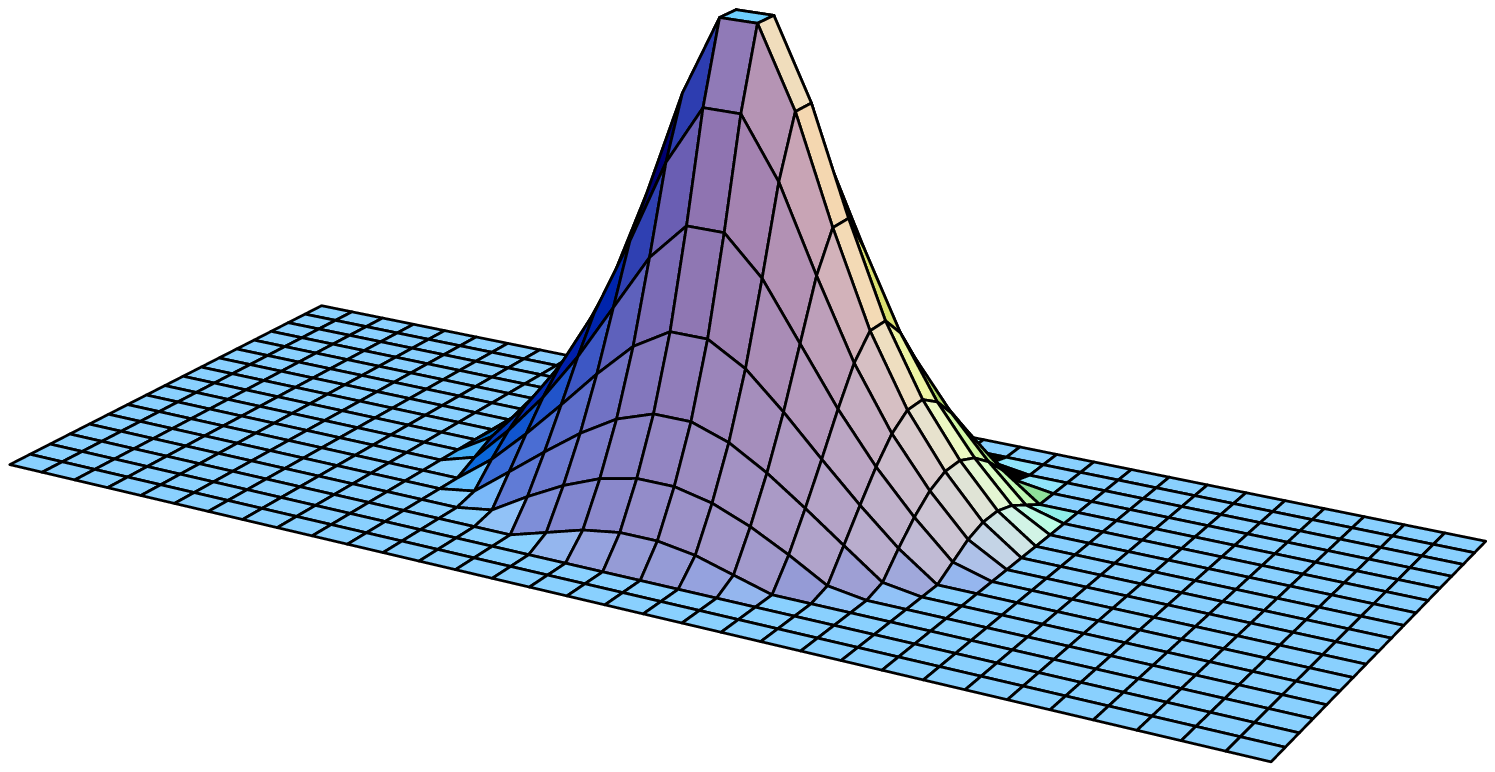}
\caption{Profiles for calorons at $\omega\!=\!0$, 0.125, 0.25 (from left to 
right) with $\rho=\cT=1$. The axis connecting the lumps, separated 
by a distance $\pi$ (for $\omega\neq0$), corresponds to the direction of 
$\hat a$. The other direction indicates the distance to this axis, 
making use of the axial symmetry of the solutions. The mass ratio of the 
two lumps is approximately $\omega/\bar\omega$, i.e. zero (no second lump), 
a third and one (equal masses), for the respective values of $\omega$.
Vertically is plotted the action density at $x_0=0$, on equal logarithmic 
scales for all profiles. They were cut off at an action density below $1/e$.}
\end{figure}

These aspects can be readily retrieved by inspecting \refeq{solution}, and in 
particular \refeq{acdensres}, for the limit of large $\rho$ and realising that 
$r$ and $s$ are the centre off mass radii of the constituent monopoles. If 
$\rho$ is small, the caloron is best described in terms of the instanton 
picture, whereas for large $\rho$ the two-monopole picture is more appropriate. 
For the constituent picture of oppositely charged BPS monopoles to be correct, 
the field has to behave like a magnetic (and electric) dipole at large 
distances. Indeed one easily finds that the field strength decays as 
$1/|\vec x|^3$ for distances much larger than $\pi\rho^2/\cT$.
Note that for $\omega=0$ we have the standard Harrington-Shepard 
caloron~\cite{HarShe}, which in the limit of large $\rho$ was already shown by 
Rossi~\cite{Ros} to become the standard BPS monopole (after a singular gauge 
transformation).

Interpreting the Nahm data (\refeq{nahmdata}) as the juxtaposition of two
sub-intervals of lengths $2\omega$ and $2\bar\omega$ respectively, with 
constant Nahm connections $\hat A(z)$, leads to a more indirect way of 
understanding the composite nature of the caloron. Indeed, $\hat A=0$ gives the
standard BPS-monopole, adding a constant merely translates the solution in 
space. Thus, each interval gives rise to a BPS monopole on $\re^3\times S^1$, 
and we can in a good approximation add the two connections corresponding to 
the two sub-intervals. The $\rho^2$ dependence of $\hat A(z)$ explains the 
large separation of the constituent lumps for large $\rho$. As the lengths of 
the intervals are given by the asymptotic Higgs vacuum expectation value of 
the corresponding monopoles, the mass ratio $\bar\omega/\omega$ of the lumps 
is easily explained by noting that the mass of a BPS monopole is proportional 
to the Higgs vacuum expectation value. The above interpretation underlies the
approach taken in ref.~\cite{LeeLu}. The expression for the gauge potential 
given there is precisely the sum alluded to above, plus gauge-like terms and 
gauge transformations required for glueing them together.

\section{The moduli space}
The moduli space $\cM$ of the caloron solutions has as its coordinates 
$\xi$ and $\zeta$. We should include the global gauge degrees of freedom 
($SU(2)$ for trivial and $U(1)$ for non-trivial holonomies), so as to
make the solution space hyperK\"ahler. The moduli space of these so-called
framed calorons is a product of the base manifold $\re^3 \times S^1$, 
parametrised by $\xi$ and the (Taub-NUT) space parametrised by $\zeta$, forming 
the non-trivial part of the moduli space and describing the relative 
coordinates of the two constituent monopoles, quite similar to that for 
the two-monopole moduli space~\cite{AtiHit}. It should be noted that
$\zeta\rightarrow-\zeta$, corresponding to the center of the $SU(2)$, 
leaves $A_\mu(x)$ invariant, such that we have to mod-out this symmetry
to obtain the space of framed calorons.

The metric on this space is given by the Riemannian metric on the gauge theory
configuration space, restricted to the space of solutions. The metric is then 
given by ($g_M^{\mu\nu}=\delta^{\mu\nu}$ being the flat metric on $M={\re^3
\times S^1}$)
\beq
g_\cM(Z,Z')=\int_Mg_M^{\mu\nu}\tr\left(Z^\dagger_\mu(x)Z'_\nu(x)
\right).\label{eq:metric}
\eeq
with $Z_\mu$ and $Z^\prime_\mu$ two vectors tangent to the space of caloron 
solutions. The gauge structure requires them to be transverse to gauge 
deformations, thereby satisfying the background (or Coulomb) gauge condition
\beq
D^\mu\!(A)Z_\mu=0.\label{eq:cougau}
\eeq
The requirement that $A_\mu+Z_\mu$ is \sd\ leads to the so-called
deformation equations
\beq
D_{[\mu}Z_{\nu]}=\half\epsilon_{\mu\nu\alpha\beta}D_{[\alpha}Z_{\beta]},
\label{eq:defeqn}
\eeq
and in the algebraic gauge we have to require in addition (see 
\refeq{permodgau})
\beq
Z_\mu(x+1)=e^{2\pi i\vec \omega\cdot\vec\tau}Z_\mu(x)
e^{-2\pi i\vec\omega\cdot\vec\tau}.\label{eq:perconzm}
\eeq

The tangent vectors (zero modes) can be found by varying the caloron
solution with respect to the coordinates $\xi $ and $\zeta$, which will
automatically satisfy the deformation equation \refeq{defeqn} and 
periodicity \refeq{perconzm}, but generally one has to apply an infinitesimal 
gauge transformation $\Phi$, compatible with \refeq{perconzm}, to transform
to the Coulomb gauge, \refeq{cougau}. Hence,
\beq
Z^r_\mu=\delta_r A_\mu+D_\mu\Phi^r,
\eeq
where the label $r$ indicates the parameters (or coordinates) of the moduli 
space. For the metric (\refeq{metric}) to exist, zero modes should of course 
be normalisable.

For instantons on $\re^4$, the zero modes can be determined within the ADHM
formalism~\cite{Osb}. Thus one can calculate the metric in terms of the ADHM 
data. This reflects the fact that the Nahm transformation (and the ADHM 
construction) is a hyperK\"ahler isometry~\cite{DonKro,BraBaa}. 
We have
\beq
\delta\Delta=\left(\bea{c}\delta \lambda\\\delta B \eea\right)\equiv
\left(\bea{c}c\\Y\eea\right)=C,
\eeq
and for the calorons, in addition periodicity (\refeq{perconzm}) requires 
\beq
Y_{m,n}=Y_{m-1, n-1},\quad c_{m+1}=e^{2\pi i\vec\omega\cdot\vec\tau}c_m.
\label{eq:perconazm}
\eeq
In terms of the deformation $C$ of the ADHM data, the zero mode reads~\cite{Osb}
\beq
Z_\mu=v^\dagger(x)C\bsigma_\mu f_xu(x)\phi^{-\half}(x)-
\phi^{-\half}(x)u^\dagger(x)f_x\sigma_\mu C^\dagger v(x).\label{eq:zerom}
\eeq
and
\beq
D^\mu Z_\mu(x)=\phi^{-1}u^\dagger(x) f_x\sigma_\mu\left(C^\dagger\Delta(x)-
\Delta^\dagger(x) C\right)\bar\sigma_\mu f_xu(x).
\eeq
(Note that for $W=W_\mu\sigma_\mu$, $\sigma_\mu W\bar\sigma_\mu=4W_0=2\tr W$.)
Combining the deformation of the quadratic ADHM constraint (as the $\myIm$ 
part) with the Coulomb gauge condition (as the $\myRe$ part), imposes
\beq
(\Delta^\dagger(x)C)=(\Delta^\dagger(x)C)^t.\label{eq:symcon}
\eeq
To satisfy the Coulomb gauge condition, the $\myRe$ part of this equation
being equivalent to
\beq
\tr(\Delta^\dagger(x)C-C^\dagger\Delta(x))=0,\label{eq:defeqadhm}
\eeq
one has only the $T$ invariance of \refeq{adhmsym} available (i.e. the $U(1)$ 
gauge invariance of $\hat A$), since a global gauge rotation would distort the 
framing. We write $T\in O(k)$ as $T=\exp(-\delta X)$, with $\delta X^t=-\delta 
X$.  Like $T$, also $\delta X$ has to satisfy $\delta X_{m,n}=\delta X_{m-n}$,
and can be interpreted as the Fourier coefficients of a gauge function 
$\delta\hat g(z)$ on the circle. We now write the zero modes $C$ in ADHM 
language as a variation of $\Delta$ plus a compensating gauge transformation,
\beq
C=\delta\Delta+\delta_X\Delta,\quad\delta_X\Delta=\left(\bea{c}
\lambda\delta X\\ \left[B,\delta X\right]\eea\right).\label{eq:var}
\eeq
Inserting this in \refeq{defeqadhm} we find
\beq
\tr\left((\delta\Delta^\dagger)\Delta-\Delta^\dagger\delta\Delta-2
\delta X\Lambda+\left[B^\dagger,\delta X\right]B-B^\dagger\left[B,
\delta X\right]\right)=0,\label{eq:comgau}
\eeq
where we used that $\left[\Lambda,\delta X\right]=0$, since the gauge symmetry 
(described by $\delta X$) is abelian. After Fourier transformation, with 
$\delta(z-z')\delta\hat X(z)=\sum_{m,n}X_{m,n}(x)e^{2\pi i(mz-nz')}$, this 
equation reads
\beq
-\frac{1}{4\pi^2}\frac{d^2\delta\hat X(z)}{dz^2}+|\zeta|^2(\delta(z-
\omega)+\delta(z+\omega))\delta\hat X(z)=\frac{i}{2}\rho^2
(\delta(z-\omega)-\delta(z+\omega))\hat\omega\cdot\vec\Sigma,
\eeq
using that with the help of the Maurer-Cartan one-forms, 
\refeq{mcartan}, we can write
\beq
\tr\left((\delta\zeta\bar\zeta-\zeta\delta\bar\zeta)\hat\omega\!\cdot\!\vec
\sigma\right)=4\hat\omega^a\eta^a_{\mu\nu}\zeta_\mu\delta\zeta_\nu=
2\rho^2\hat\omega\cdot\vec\Sigma.
\eeq
The solution to the differential equation for $\delta\hat X(z)$ gives the
infinitesimal gauge transformations needed to go to Coulomb gauge. One finds
\beq
\delta\hat X(z)=-2\pi^2i\rho^2\hat\omega\cdot\vec\Sigma(1+8\pi^2
\omega\bar\omega\rho^2)^{-1}\int_0^z\Theta_\omega(z')dz'\label{eq:gtraf}
\eeq
which is a zig-zag wave (see fig.~5), vanishing at $2z\in\zahlen$ and taking 
its extremal values at $z=\pm\omega$,
\beq
\delta\hat X_\omega\equiv\pm\delta\hat X(\pm\omega)=-4i\pi^2\omega\bar\omega
\rho^2\hat\omega\cdot\vec\Sigma(1+8\pi^2\omega\bar\omega\rho^2)^{-1},
\eeq
Note that for the variations with respect to the caloron position, $\xi$, no 
compensating gauge transformation is needed. 

\begin{figure}[htb]
\vspace{3.6cm}
\includegraphics{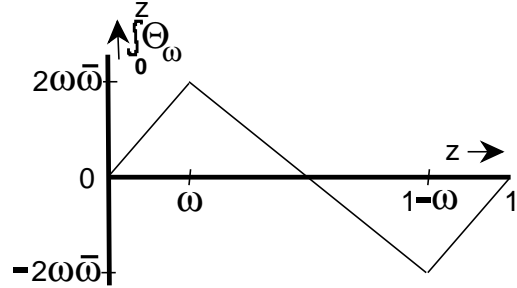}
\caption{The function $\int^z_0\Theta_\omega(z')dz'$.}
\end{figure}

In order to evaluate the metric, \refeq{metric}, it is sufficient to compute
$g_\cM(Z,Z)$ for $Z$ related to an arbitrary deformation of the moduli 
parameters, as determined by $C$ in \refeq{var} and \refeq{gtraf}. For this 
we employ the following relation due to Osborn~\cite{Osb}
\beqa
&&\tr(Z^\dagger_\mu(x) Z^{\phantom{\dagger}}_\mu(x))=-\half\partial^2\,\tr\,
\Tr\,\left(C^\dagger(2-\Delta(x)f_x\Delta^\dagger(x))Cf_x\right)\\
&&\hskip2.85cm=-\half\partial^2\,\tr\,\Tr\,\left(2(c^\dagger c+Y^\dagger Y)f_x
-(c^\dagger\lambda+Y^\dagger B(x))f_x(\lambda^\dagger c+B^\dagger(x) Y)f_x
\right),\nonumber
\eeqa
which is derived from \refeq{zerom}, making use of \refeq{symcon}. We introduce
the Fourier transforms $\hat c(z)\equiv\sum_m\exp(2\pi imz)c_m$ and $\hat Y(z)$,
with $\delta(z-z')\hat Y(z)=\sum_{m,n}e^{2\pi i(mz-n z')}Y_{m,n}$, such that
\beqa
\hat c(z)&=&\delta\hat\lambda(z)+\hat\lambda(z)\delta\hat X(z)
=P_+\delta(z-\omega)(\delta\zeta+\zeta\delta\hat X_\omega)\label{eq:maux}
+P_-\delta(z+\omega)(\delta\zeta-\zeta\delta\hat X_\omega),\nonumber\\
\quad\hat Y(z)&=&\frac{1}{2\pi i}\left(\delta\hat A(z)+\ddz\delta\hat X(z)
\right)\nonumber\\&=&\delta\xi+\pi\left(\delta\bar\zeta\hat\omega\cdot\vec\sigma
\zeta+\bar\zeta\hat\omega\cdot\vec\sigma\delta\zeta-\rho^2\hat\omega\cdot
\vec\Sigma(1+8\pi^2\omega\bar\omega|\zeta|^2)^{-1}\right)\Theta_\omega(z).
\eeqa
We may use the special structure of $\hat c(z)$ and $\hat\lambda(z)$ 
(\refeq{fourtr}) - formed out of the combinations $\delta(z\mp\omega)P_\pm$, 
with $P_\pm^2=P_\pm$ and $P_\pm P_\mp=0$ - to deduce
\beq
\hat c^\dagger(z)\hat c(z')\!=\!\delta(z-z')\hat c^\dagger(z)\!<\!\hat c\!>,
\quad\hat c^\dagger(z)\hat\lambda(z')\!=\!\delta(z-z')\hat c^\dagger(z)
\!<\!\hat\lambda\!>,\quad <\!h\!>\equiv\!\!\int_{S^1}\!h(z)dz.\label{eq:simp}
\eeq
For calorons, this allows us to turn Osborn's equation into
\beqa
&&\tr(Z^\dagger_\mu(x) Z^{\phantom{\dagger}}_\mu(x))=-\partial^2\int_{S_1}
dz~\tr\left(\left[\hat Y^\dagger(z)\hat Y(z)+\hat c^\dagger(z)<\!\hat c\!>
\right]\hat f_x(z,z)\right)\label{eq:osb}\\&&\hskip3cm+\half\partial^2
\int_{S^1}dzdz'~\tr\left([\cC(z)+\cY_x(z)]\hat f_x(z,z')[\cY_x^\dagger(z')+
\cC^\dagger(z')]\hat f_x(z',z)\right),\nonumber
\eeqa
with
\beq
\cY_x(z)=(2\pi i)^{-1}\hat Y^\dagger(z)\hat D_x(z),\quad \cC(z)=
\hat c^\dagger(z)<\!\hat\lambda\!>.
\eeq

When integrating over space-time, the $\partial_0^2$ part does not contribute
due to periodicity. The integral is therefore reduced to a boundary term at 
spatial infinity, $|\vec x|\rightarrow\infty$. In this limit $\rho^2$ can be
neglected and the two radii $r$ and $s$ in \refeq{radii} become equal. In
particular in this limit the potential in \refeq{wa} equals $|\vec x|^2$, 
independent of $z$.  From this one concludes that asymptotically 
$\hat f_x(z,z')$ becomes a function of $z-z'$ and therefore that
$\hat D_x(z)\hat f_x(z,z')\hat D^\dagger_x(z')=4\pi^2\delta(z-z')+\cO(r^{-1})$. 
This can also be deduced from the asymptotic form of $\hat f_x(z,z')$ in 
\refeq{fas}, which implies $\hat f_x(z,z)=\pi/r+\cO(r^{-2})$. Thus
$\int_{S^1}dz'~\tr[\cY_x(z)\hat f_x(z,z')\cY_x(z')\hat f_x(z,z')]$ 
$=\tr[\hat Y^\dagger(z)\hat Y(z)\hat f_x(z,z)]$, to be combined with the first 
term in \refeq{osb}. Using \refeq{simp}, we also have $\int_{S^1}dz'~\tr[\cC(z)
\hat f_x(z,z')\cC^\dagger(z')\hat f_x(z,z')]\!=$ $\tr[<\!\hat c^\dagger\!>
<\!\hat\lambda\!><\!\hat\lambda^\dagger\!>\hat c(z)\hat f^2_x(z,z)]=
\cO(r^{-2})$, and $2\pi i\int_{S^1}dz'~\tr[\cY(z)\hat f_x(z,z')\cC^\dagger(z')
\hat f_x(z,z')]\!=$ $\sum_{s=\pm}\tr[\hat Y^\dagger(z)D_x(z)f_x(z,s\omega)
<\!\hat\lambda^\dagger\!><\!\hat c\!>P_sf_x(s\omega,z)]$, which after 
integration over $z$ is $\cO(r^{-2})$. Only those terms that are $\cO(r^{-1})$ 
will contribute and we obtain the following remarkably simple result
\beq
g_\cM(Z,Z)=2\pi^2\tr(<\hat Y^\dagger\hat Y>+2<\hat c^\dagger><\hat c>).
\label{eq:metrics}
\eeq
Inserting \refeq{maux} gives the metric (we put $\hat\omega = \hat e_3$)
\beq
ds^2=2\pi^2\left\{2d\xi_\mu d\xi^\mu\!+\!(1\!+\!8\pi^2\omega\bar\omega\rho^2)
\left(4d\rho^2\!+\!\rho^2\left( \Sigma^2_1\!+\!\Sigma^2_2\right)\right)\!+\!
\rho^2(1\!+\!8\pi^2\omega\bar\omega\rho^2)^{-1}\Sigma^2_3\right\}.
\eeq
The first part describes the flat metric of the base manifold $\re^3 \times
S^1$, the remainder forms the non-trivial part of the metric. They separate 
because $\int\Theta_\omega(z)dz=0$, see \refeq{theta}. 

We introduce a ``radial" coordinate $X$ and ``mass" parameter $M$ 
\beq
X^2=8\pi^2\rho^2,\quad M^{-2}={16\omega\bar\omega},
\eeq
and rewrite the non-trivial part of the moduli space metric as
\beq
ds^2=\left(1+\frac{X^2}{16M^2}\right)\left(dX^2+\quarter X^2(\Sigma^2_1+
\Sigma^2_2)\right)+\quarter X^2\left(1+\frac{X^2}{16M^2}\right)^{-1}
\Sigma^2_3.\label{eq:taubnut1}
\eeq
This metric is the Taub-NUT metric~\cite{Nut} as given in~\cite{Haw1}. It is a 
self-dual Einstein manifold~\cite{Haw2,SdEi} and is 
hyperK\"ahler~\cite{AtiHit,HKLR}. The latter property is inherited from the 
hyperK\"ahler structure of the base manifold $\re^3\times S^1$, preserved 
by the \sd ity equations~\cite{DonKro}. Therefore, the $\su2$ moduli space 
for calorons becomes
\beq
\cM_{\rm framed}=(\re^3 \times S^1)\times{\rm Taub}\!-\!{\rm NUT}/Z_2.
\eeq
Note that $Z_2$ corresponds with $\zeta=q=\pm1$, i.e. the center of the
\su2\ gauge group. With $\zeta\rightarrow-\zeta$ leaving the gauge fields
unchanged, this gives rise to an orbifold singularity (at $\zeta\!=\!0$) and 
$(\re^3\!\times\!S^1)\!\times\!{\rm Taub}\!-\!{\rm NUT}$ is a double cover 
of $\cM_{\rm framed}$. 

For small $\rho$ or $\omega$, when $X^2/M^2\rightarrow0$, the metric 
becomes that of $\re^4$, since $\quarter(\Sigma^2_1+\!\Sigma_2^2+\!\Sigma_3^2)$
(see \refeq{mcartan}) is the metric on the unit three-sphere. With $\rho
\rightarrow 0$ corresponding to $\cT\rightarrow\infty$, this describes the 
moduli space of a charge one instanton on $\re^4$, whereas for $\omega=0$ we 
have the standard Harrington-Shepard caloron moduli space. In both cases 
this is parametrised by the scale and $\su2$ group orientation (to make 
the moduli space hyperK\"ahler) and $\cM_{\rm framed}=\re^4\!\times\!
\re^4/Z_2$, see ref.~\cite{DonKro,GauHar}.

For large $\rho$ (i.e. large $X$), or equivalently for $\cT\rightarrow0$, 
Taub-NUT space is a squashed $S^3$, that is $S^2\times S^1$, with $S^1$
a non-trivial (Hopf) fibration over $S^2$. This is best studied by introducing 
a radial coordinate $R$ through $X^2=8MR$, which brings the Taub-NUT metric 
to the form~\cite{Haw2}
\beq
ds^2=\left(1+\frac{2 M}{R}\right)(dR^2+R^2(d\theta^2+
\sin^2\theta d\phi^2))+4M^2\left(1+\frac{2 M}{R}\right)^{-1}
(d\Upsilon+\cos\theta d\phi)^2,\label{eq:kklein}
\eeq
also familiar from the (spatial part of the) Kaluza-Klein monopole 
solution~\cite{SoGrPe}. Since $\Upsilon\in[0,4\pi]$ we read-off from the 
asymptotic form of the metric that the compactification radius equals $4M$. At 
large $\rho$, $\cM_{\rm framed}$ is therefore of the form $(\re^3\!\times\!S^1)
\!\times\!(\re^3\!\times'\!S^1)/Z_2$. It is natural to view this as the product 
space of two single BPS monopole moduli spaces. The first $\re^3$ represents 
the centre of mass and the second the relative coordinates. The first $S^1$ 
corresponds to $x_0$ and can be seen as a global $U(1)$. The other gives the 
relative $U(1)$ orientation on which the $Z_2$ acts. This $Z_2$ does {\em not} 
act on the positions, as the monopoles have in general different masses and 
are hence not identical objects. A similar description is valid for the $\su2$
two-monopole moduli space, see ref.~\cite{GibMan,AtiHit}. At {\em large} 
separations its metric is precisely of the Taub-NUT form, but with a {\em 
negative} mass parameter $M$, as was shown by Manton~\cite{Man}, using the 
asymptotic form of the interactions. The complete metric was constructed by 
Atiyah and Hitchin (see ref.~\cite{AtiHit}) in terms of elliptic integrals, 
using its symmetries and hyperK\"ahler structure.

Finally it is interesting to note that the moduli space of an $SU(3)$ monopole 
with maximal symmetry breaking to $U(1)\!\times\!U(1)$ and charges $(1,1)$ 
(see ref.~\cite{Con,su3}) is Taub-NUT with a positive mass parameter, as for 
the caloron. This is not surprising, as its Nahm data are precisely those of 
the caloron. For an $SU(3)$ monopole with maximal symmetry breaking 
there are two sub-intervals $[\mu_1,\mu_2]$ and $[\mu_2, \mu_3]$. On each 
sub-interval, $\hat A$ has dimension one and is therefore constant, 
due to the Nahm equations. The matching at $\mu_2$, implied by the delta 
function singularity, is of the correct form (defining $\half\rho^2$), and 
implies we can identify $\mu_1$ with $\mu_3$ (defining $\cT$) with correct
matching as well. As the metric can be formulated in terms of the Nahm data,
we would indeed expect the metric to be the same.

\section{Discussion}

The interpretation of the ADHM data for a periodic instanton as the Fourier 
coefficients of the Nahm transformed Weyl operator extends naturally to higher
charges and other gauge groups~\cite{LeeYi}. For charge one calorons in $SU(N)$ 
the determination of the Green's function remains a problem of quantum mechanics
on the circle with a piecewise constant potential (on $N$ sub-intervals, 
separated by delta functions). Also our formalism to compute the metric on the 
moduli space can be generalised. Due to the relation with the ADHM approach, 
one may wonder~\cite{LeYi} if there is some advantage in obtaining monopole 
solutions from the calorons by sending certain scales to infinity - in the 
limit of which the solution becomes static and constituents separate. For 
higher charges the Nahm bundle is no longer abelian and the construction is 
more complicated. For generalisations to further compactifications, \eg\ 
$\re^2\times T^2$ and $\re\times T^3$ (see ref.~\cite{BaaT3R}), note that the 
't Hooft ansatz~\cite{JacNohReb} diverges when summing over more than one 
direction. This will correspond to all holonomies trivial and one may well 
have no solutions in that case. A dramatic particular example of a 
non-existence proof for charge one instantons is $T^4$, see ref.~\cite{BraBaa},
a situation where indeed an existence proof of Taubes~\cite{Tau} does not hold.

\begin{figure}[htb]
\vspace{3.3cm}
\includegraphics{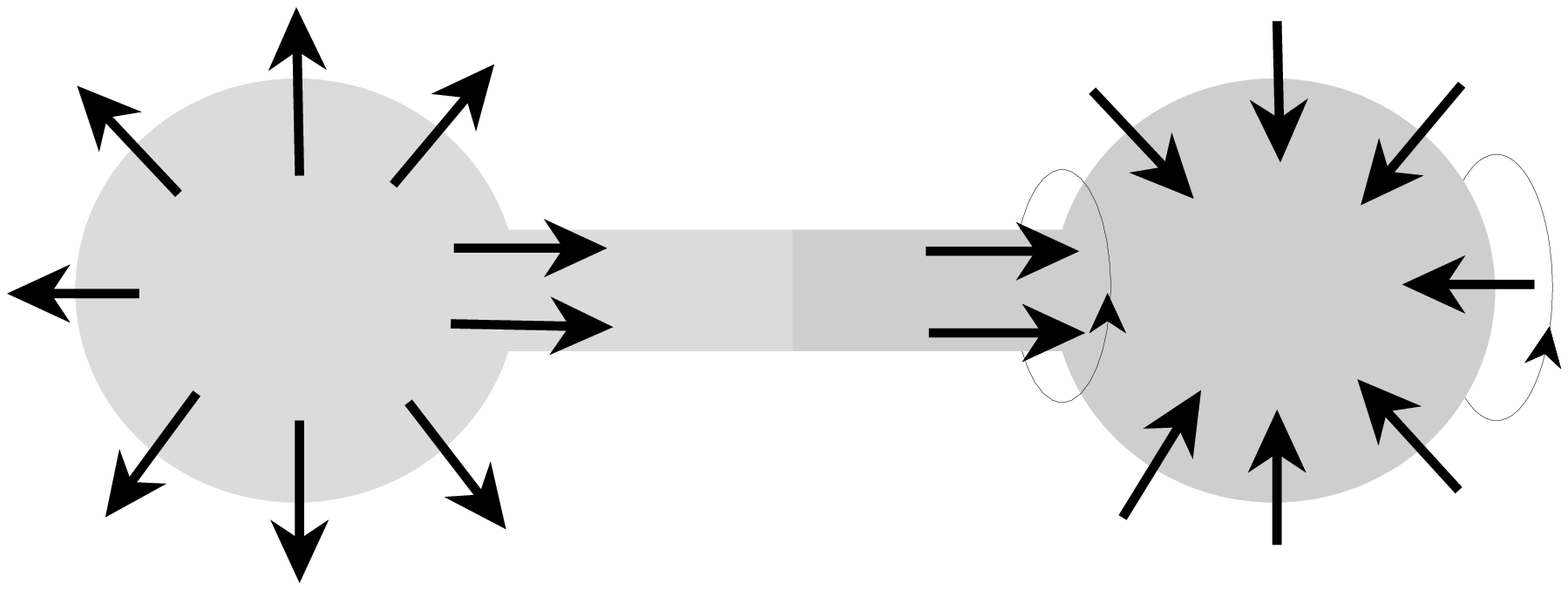}
\includegraphics{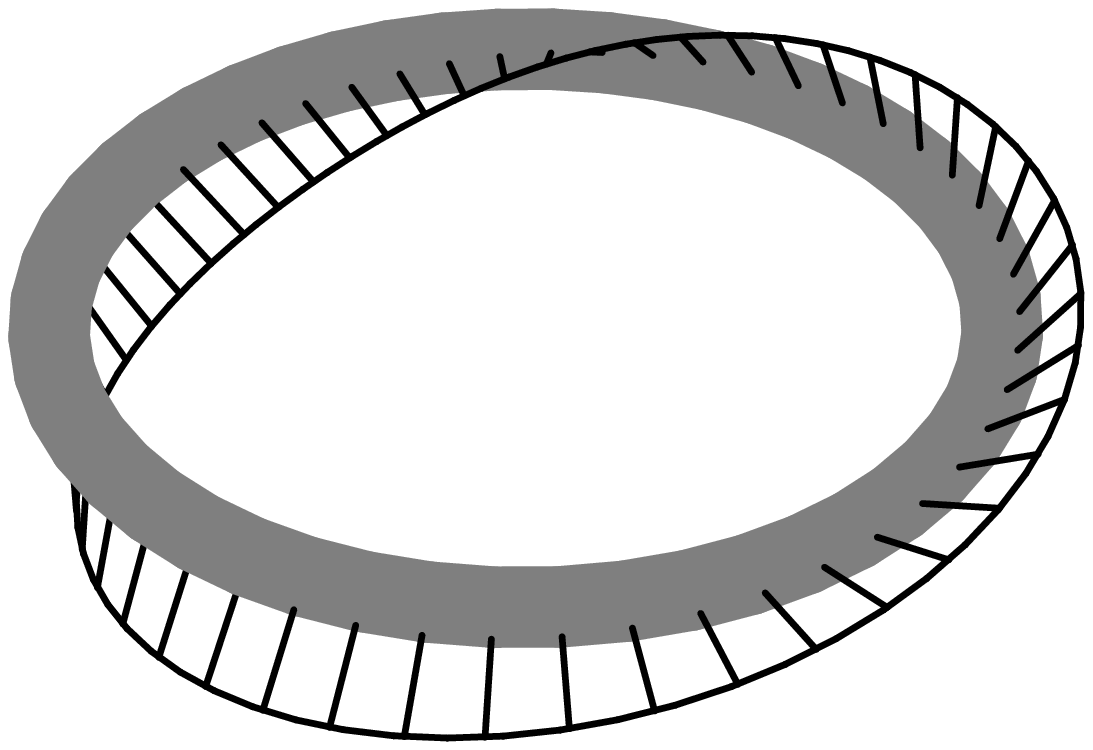}
\caption{The non-contractible loop is constructed from two oppositely charged
monopoles by rotating one of them, as indicated on the left. On the right
is a closed monopole line, rotating its frame when completing the circle.}
\end{figure}

We now recall briefly Taubes's arguments for building gauge fields with 
topological charge one out of monopole fields~\cite{Taubes,TauCMP}. Although his
construction was within the standard model with a genuine Higgs field, the same
argument applies to the caloron case, using $A_0$ as the Higgs field. As we saw
in sect.~2 (eqs.~(\ref{eq:rotdiag},\ref{eq:asa0})), non-trivial $\su2$ 
monopole fields can be classified by the winding number $k_1=-k_2$ of maps from
$S^2$ to $SU(2)/U(1)\sim S^2$. We consider at this point configurations at a 
{\em fixed} time $t$, $\Psi=\{A_\mu(\vec x)\}$. In the sector where the net 
winding vanishes, we study a one-parameter family of configurations, $\Psi_t=
\{A_\mu(\vec x,t)\}$ (the parameter can, but need not, be seen as the time $t$).
When this configuration is made out of monopoles with opposite charges, in a 
suitable gauge the isospin orientations behave as shown in fig.~6, sufficiently
far from the core of both monopoles. We note that the arrows match in the 
``throat'' of the configuration. This remains true if we rotate {\em only one} 
of the monopoles around the axis of the throat. Clearly, the net magnetic 
winding remains zero, but the fields of two monopoles will no longer cancel 
when brought together, despite the fact that the long range abelian components 
do cancel. The non-contractible loop is now constructed by letting $t$ affect
a {\em full} rotation.

Taubes describes this by creating a monopole anti-monopole pair, bringing 
them far apart, rotating one of them over a full rotation and finally bringing 
them together to annihilate. The four dimensional configuration constructed 
this way is topologically non-trivial. Since an anti-monopole travelling 
forward in time is a monopole travelling backwards in time, we can describe 
the same as a closed monopole line (or loop). It represents a topologically 
non-trivial configuration when the monopole makes a full rotation while moving 
along the closed monopole line (see fig.~6). The non-trivial topology discussed
by Taubes~\cite{Taubes} ($\pi_1(M_0(S^2;S^2))=\zahlen$) is just the Hopf 
fibration, except that now it is more natural to see $S^1$ as the base manifold
and $S^2$ as the fibre, which rotates (twists) while moving along the circle 
formed by the closed monopole line . The only topological invariant available 
to characterise this homotopy type is precisely the Pontryagin index. 

We mentioned that the short range components of the fields can not be fully
cancelled due to the non-trivial ``twist'' along the monopole line, so they 
have to be  responsible for the Pontryagin index. Indeed, in the computation 
of the total topological charge of our configurations as the integral over 
$\partial_\mu^2\partial_\nu^2\log\psi$ (\refeq{pontr}), the massive component 
of the field gives rise to 
\beq
\psi=2e^{4\pi(r\bar\omega+s\omega)}(1+\cO(|\vec x|^{-1}))=2e^{2\pi|\vec x|}(1+
\cO(|\vec x|^{-1})),\quad\partial_\mu^2\log\psi=4\pi/|\vec x|+
\cO(|\vec x|^{-4}),
\eeq
and thus yields the surviving boundary term, but at the same time does not
contribute to the action density, since $\partial_\mu^2|\vec x|^{-1}=0$.

We now inspect more closely the monopole content of our calorons. For this we 
choose $\rho/\cT$ large, such that the monopoles are well separated and static.
We have two world lines of monopoles running in opposite directions (due to 
their opposite charges), closed due to the periodic boundary conditions. At 
smaller separations the solutions are far from static, with the attractive 
force driving the constituents together, after which they annihilate. In that 
case the world lines form a single closed monopole line, as mentioned above. 
It should be noted though, that for small $\rho/\cT$ the constituents become 
rather extended. Nevertheless, such closed monopole lines are characterised 
by rotation of the local monopole field over precisely one full rotation when 
completing the circle, since our solution has unit topological charge. It
prevents the field from decaying to the trivial configuration. It is this 
``twist'' that provides the closed monopole line its stability. 

Before continuing, we observe that the calorons are given in a singular gauge,
as is usual for the ADHM construction. The function $\hat\psi$ (\refeq{psihat})
has an isolated zero at $x=0$. This can be traced to the zero-mode in $B-x$, 
responsible for the non-trivial topology of the solution. This singularity is 
easily seen to be removed by a gauge transformation, locally of the form 
$x/|x|$ (viewing $x$ as a quaternion). We now assume that $\omega\bar\omega\neq
0$ and consider the region outside the core of both monopoles, i.e. $r\bar\omega
>1$ and $s\omega>1$. In this region
\beqa
\phi&=&\frac{r+s+\pi\rho^2}{r+s-\pi\rho^2}+\cO(e^{-8\pi\,{\rm min}(s
\omega,r\bar\omega)}),\label{eq:abproj}\\
\tilde\chi&=&\frac{4\pi\rho^2}{(r+s+\pi\rho^2)^2}\left\{re^{-4\pi r\bar\omega}
e^{-2\pi ix_0}+s e^{-4\pi s\omega}\right\}\left(1+\cO(e^{-8\pi\,{\rm min}(s
\omega,r\bar\omega)})\right).\nonumber
\eeqa
Substituting this in \refeq{solutionpg} we find the solution to be
time independent and abelian, up to exponential correction
\beqa
&&A_0=\frac{i}{2}\tau_3\partial_3\log\phi,\quad
A_k=\frac{i}{2}\tau_3\epsilon_{kj3}\partial_j\log\phi,\\
&&E_k=F_{k0}=\frac{i}{2}\tau_3\partial_k\partial_3\log\phi,\quad
B_k=\epsilon_{kij}\partial_iA_j=\frac{i}{2}\tau_3\left(\partial_k\partial_3
\log\phi-\delta_{k3}\partial_j^2\log\phi\right).\nonumber
\eeqa
For convenience we rotate $\hat a$ to $\hat e_3$. Self-duality, $\vec E\!=\!
\vec B$, requires $\log\phi$ to be harmonic. We first note that when {\em 
neglecting} the exponential corrections, $\phi^{-1}$ vanishes on the interval 
$-2\pi\rho^2\omega\leq x_3\leq2\pi\rho^2\bar\omega$ at $x_1\!=\!x_2\!=\!0$
(we denote the characteristic function on this interval by $\chi_\omega(x_3)$).
A careful analysis reveals $\partial_j^2\log\phi\!=\!-4\pi\delta(x_1)
\delta(x_2)\chi_\omega(x_3)$ (that it vanishes away from the zeros of 
$\phi^{-1}$ follows by a direct computation). The term $-\frac{i}{2}\tau_3
\delta_{k3}\partial_j^2\log\phi$ in the expression for the magnetic field 
corresponds precisely to the Dirac string singularity, carrying the return 
flux. One finds that $\partial_kE_k\!=\!\partial_kB_k\!=\!\frac{i}{2}\tau_3
\partial_3\partial_j^2\log\phi\!=\!2\pi i\tau_3(\delta_3(\vec s)\!-\!
\delta_3(\vec r))$, when ignoring this return flux, which in the full theory 
is absent~\cite{Mon} (indeed as noted before $\phi^{-1}$ has only an isolated 
zero at $x=0$, corresponding to a gauge singularity).

Finally, to confirm our expectations it remains to identify the rotation of 
one of the monopoles so as to guarantee the topologically non-trivial nature 
of the configuration. Inspecting the behaviour in the core region of the 
monopoles,  described by $\tilde\chi$ in \refeq{abproj}, gives the following 
factorisation 
\beq
\tilde\chi=\chi^{(1)}(r)+e^{-2\pi ix_0}\chi^{(2)}(s).
\eeq
While one of the monopoles has a static core, the other has a time
dependent phase rotation - equivalent to a (gauge) rotation - precisely
of the type required to form a non-contractible loop, as the phase makes a 
full rotation when closing by the periodic boundary conditions in the time 
direction.

Although interpreting $A_0$ as the Higgs field allows one to introduce
monopoles in pure gauge theory, there are some subtle differences. In the 
static limit the BPS equations imply $F_{i0}=D_iA_0$ and we would be tempted 
to call the solution a dyon. In the Higgs model dyons are constructed by 
taking $A_0$ proportional to the Higgs field $\Phi$~\cite{BPS,JulZee}. By a 
time dependent gauge transformation $A_0$ can be gauged to zero. This gauge 
transformation is generated by $A_0$, precisely the unbroken generator, as 
$A_0$ is proportional to the Higgs field. The resulting electric field is 
now given by $\partial_0A_i$ and is {\em not} quantised. In the abelian Higgs 
model $B_i=D_i\Phi$ and $E_i=-\partial_0A_i$. In pure gauge theory it makes,
however, no sense to separate $D_i\Phi=D_iA_0$ from $\partial_0A_i$. Gauge 
invariance requires that they occur in the combination $F_{i0}=
D_iA_0-\partial_0A_i$. The electric field is necessarily fixed and quantised 
as soon as we interpret $A_0$ as the Higgs field. Nevertheless, we {\em can} 
consider dyons also in pure gauge theories, but for this we have to add a 
term proportional to $\theta F\tilde F$ to the lagrangian~\cite{Wit}. The 
electric charge is now proportional to $\theta$, so no longer quantised. But, 
unlike in the Higgs model, it is the same for all monopoles. 

We note that in 
the Higgs model the construction of the non-contractible loop generates an 
electric charge due to the (gauge) rotation along the closed monopole line, 
when interpreting the loop parameter as time. The electric charge is 
proportional to the rate of rotation and can vary along the monopole line. 
However, integrated along a closed monopole line the charge is fixed and
proportional to the number of rotations, which hence plays the role of a 
winding number. In pure gauge theory this winding can not be read off from 
the long range field components, but for both cases the fields in the core 
are responsible for the Pontryagin number (an abelian field can not contribute 
to this topological charge).

There is a natural context in which the analogy with the Higgs model is more 
precise. For this we have to add time as a fifth dimension, such that four
dimensional space is compactified on a circle. In the limit of zero 
compactification radius ($\cT\rightarrow0$) our solutions become genuine 
monopoles and can obtain dyonic charges in the sense of Julia and 
Zee~\cite{JulZee}. It is in this context that Taub-NUT metric describes the 
scattering of oppositely charged monopoles on $\re^3\times S^1$, in exactly
the same way as the Atiyah-Hitchin metric describes the scattering of 
like-charged monopoles on $\re^3$~\cite{Manton,AtiHit}.

Monopoles appear also in the context of 't Hooft's abelian 
projection~\cite{Abproj} as (gauge) singularities. The lesson we have
learned from the above analysis is that in order to include the non-trivial
topological charge, important for fermion zero modes, breaking of the 
axial $U(1)$ symmetry~\cite{Hoo} and presumably for chiral symmetry breaking, 
one needs to keep some information on the behaviour near the core of these
monopoles. This allows one to combine the attractiveness of the dual 
superconductor picture of confinement~\cite{Super} in terms of monopole 
degrees of freedom, with the success of the instanton liquid model~\cite{Shur}.
There have been many attempts to make an effective monopole model for the
long range confining properties of QCD, see e.g. ref.~\cite{Smit}. Also there
have been many studies in lattice gauge theory, using the idea of abelian 
projection implemented by the so-called maximal abelian gauge~\cite{MaxAb},
in order to extract the monopole content of the theory. It was 
observed that the string tension is saturated by the monopole 
fields~\cite{MonDom}. More recently it was found that after abelian projection 
instantons contain closed monopole lines~\cite{MonInst}. As emphasised in 
ref.~\cite{Edin}, in the light of Taubes's construction this was to be
expected. Here we have shown in more detail how one can make fields with 
non-zero topological charge out of monopole degrees of freedom, with as example
the well defined setting of calorons with non-trivial Polyakov loop. What is 
{\em minimally} required is a frame associated to each monopole, whose 
rotation is a topological invariant for closed monopole lines. Such closed 
monopole lines can shrink, but one will be left over with what represents an 
instanton. It would be interesting to build a hybrid model based on the 
instanton liquid and monopoles, and see how successful it is in capturing the 
appropriate phenomenology.

To conclude, it is sensible to take the monopole content of instantons 
serious in the broader context sketched here. Our gauge invariant method of 
investigating the monopoles inside an instanton is somewhat destructive
(but reversible). First we heat the instanton just a little. Then we add a 
non-trivial value of the Polyakov loop at infinity, without disturbing the 
instanton significantly (true for $\cT$ sufficiently large). Now we have to 
squeeze (or heat) it hard. Out come the two constituent monopoles, in a 
direction determined by the choice we have made for the Polyakov loop at 
infinity (which does not change under heating). The new caloron solutions 
can be studied on the lattice by taking all links in the time direction,
at the spatial boundary of the lattice, equal to $U_0=\exp(2\pi i\vec\omega
\cdot\vec\tau/N_0)$ (in lattice units $\cT$ equals $N_0$). One can look 
for solutions using improved cooling~\cite{ICool} (to prevent calorons to 
disappear due to scaling violations). When interested in seeing the monopole 
constituents one may just as well take the time direction to be one lattice 
spacing ($N_0=1$). The lattice study of ref.~\cite{LauSch} hints at regular 
monopole solutions in pure gauge theory, although with a vanishing electric 
component, not what we would expect for our constituents.

\section*{Acknowledgements}

We thank Kimyeong Lee for useful correspondence and Dmitri Diakonov,
Simon Hands and Bernd Schroers for discussions. TCK was supported by a 
grant from the FOM/SWON Association for Mathematical Physics.

\section*{Appendix}
In this appendix we present the result for the Green's function $G_x=
(B^\dagger(x)B(x))^{-1}$, which after Fourier transformation satisfies the
equation
\beq
\left\{\!\!\left(\!\frac{1}{2\pi i}\ddz\!-\!x_0\right)^{\!\!2}\!\!+\!s^2
\chi_{[-\omega,\omega]}(z)\!+\!r^2\chi_{[\omega,1-\omega]}(z)\!-\!
\frac{\rho^2}{2}\hat\omega\!\cdot\!\vec\tau(\delta(z\!\!-\!\!\omega)\!-
\!\delta(z\!\!+\!\!\omega))\!\right\}\!\!\hat G_x(z,z')=\delta(z-z'),
\eeq
with $r$ and $s$ given as in \refeq{radii}. Its solution is given by
\beqa
&&\hskip-5mm \hat G_x(z,z')=\chi_{[-\omega,\omega]}(z')\left(\chi_{[-\omega,
\omega]}(z)\hat G^d_x(z,z',r,s,\omega)+\chi_{[\omega,1-\omega]}(z)\hat 
G^o_x(z,z',r,s,\omega)\right)\\&&\hskip-1mm+\chi_{[\omega,1\!-\!\omega]}(z'))
\left(\chi_{[\omega,1-\omega]}(z)\hat G^d_x(\half\!-\!z,\half\!-\!z',s,r,
\bar\omega)^*\!+\!\chi_{[-\omega,\omega]}(z)\hat G^o_x(\!z',z,r,s,\omega)^*
\right).\nonumber
\eeqa
Like for $\hat f_x$ the diagonal component $\hat G^d_x(z,z')$ is only defined 
strictly for $z,z'\in[-\omega,\omega]$ and the off-diagonal component 
$\hat G^o_x(z,z')$ only for $z\in[\omega,1\!-\!\omega]$ and $z'\in[-\omega,
\omega]$. For $z$ or $z'$ outside of these intervals, one first has to map 
back to the interval $[-\omega,1\!-\!\omega]$, using periodicity.
\beqa
&&\hskip-5mm\hat G^d_x(z,z',r,s,\omega)=e^{2\pi ix_0(z-z)}\pi(rs\hat\psi)^{-1}
\Big\{e^{-2\pi ix_0{\rm sign}(z-z')}r\sinh(2\pi s|z\!-\!z'|)\\&&\qquad+
s^{-1}\sinh(4\pi r\bar\omega)\Big[\pi s\rho^2\hat\omega\cdot\vec\tau\sinh(2
\pi s(z+z'))\!+\!\half(s^2\!-\!r^2\!+\!\pi^2\rho^4)\cosh(2\pi s(z\!+\!z'))
\nonumber\\&&\qquad+\!\half(s^2\!+\!r^2\!-\!\pi^2\rho^4)\cosh(2\pi s(2\omega
\!-\!|z\!-\!z'|))\Big]\!+\!r\cosh(4 \pi r \bar\omega)\sinh(2\pi s(2\omega
\!-\!|z\!-\!z'|))\Big\},\nonumber\\
&&\hskip-5mm\hat G^o_x(z,z',r,s,\omega)=e^{2\pi ix_0(z- z')}\pi(rs\hat\psi)^{-1}
\Big\{\pi\rho^2\hat\omega\cdot\vec\tau\sinh(2\pi r(1\!-\!z-\!\omega))
\sinh(2\pi s(z'\!+\!\omega))\nonumber\\&&\qquad+\!r\cosh(2\pi r(z\!-\!1\!+\!
\omega))\sinh(2\pi s(z'\!+\!\omega))-s\sinh(2\pi r(z\!-\!1+\!\omega))
\cosh(2\pi s(z'\!+\!\omega))\nonumber\\&&\qquad e^{-2\pi ix_0}\Big[
s\sinh(2\pi r(z\!-\!\omega))\cosh(2\pi s(z'\!-\!\omega))-r\cosh(2\pi r(z\!-\!
\omega))\sinh(2\pi s(z'\!-\!\omega))\nonumber\\&&\hskip6cm+\pi\rho^2\hat\omega
\cdot\vec\tau\sinh(2\pi r(z\!-\!\omega))\sinh(2\pi s(z'\!-\!\omega))\Big]\Big\}.
\nonumber
\eeqa
In particular,
\beqa
&&\hat G_x(\omega,-\omega)=\pi(rs\hat\psi)^{-1}e^{4\pi ix_0\omega}
\left\{e^{-2\pi ix_0}r\sinh(4\pi s\omega)+s\sinh(4\pi r\bar\omega)\right\},\\
&&\hat G_x(\pm\omega,\pm\omega)\!=\!\pi(rs\hat\psi)^{-1}\Big\{s\sinh(4\pi r\bar
\omega)\cosh(4\pi s\omega)\!+\!r\sinh(4\pi s\omega)\cosh(4\pi r\bar\omega)
\nonumber\\&&\hskip7.8cm\pm\pi\rho^2\hat\omega\!\cdot\!\vec\tau
\sinh(4\pi r\bar\omega)\sinh(4\pi s\omega)\Big\}.\nonumber
\eeqa
which can be used to verify \refeq{valnorm} as derived from \refeq{norm}, and 
\refeq{detG}.


\begin{thebibliography}{99}
\bibitem{AtiDriHitMan}M.F. Atiyah, N.J. Hitchin, V.G. Drinfeld, Yu. I. Manin,  
Phys. Lett. {\bf 65 A} (1978) 185.
\bibitem{BPS}E.B. Bogomol'ny, Yad. Fiz. {\bf 24} (1976) 861; Sov. J. Nucl. 
{\bf 24} (1976) 449; M.K. Prasad and C.M. Sommerfield, Phys. Rev. Lett. 
{\bf 35} (1975) 760.
\bibitem{Man0}N.S. Manton, Nucl. Phys. {\bf B126} (1977) 525.
\bibitem{MoOl}C. Montonen and D. Olive, Phys. Lett. {\bf 72B} (1977) 117.
\bibitem{GarMur}H. Garland and M.K. Murray, Commun. Math. Phys. {\bf 120} 
(1988) 335.
\bibitem{LeYi}K. Lee and P. Yi, Phys. Rev. {\bf D56} (1997) 3711
(hep-th/9702107).
\bibitem{NahMonCal}W. Nahm, {\it Self-dual monopoles and calorons}, in:
Lect. Notes in Physics. 201, eds. G. Denardo, e.a. (1984) p. 189.
\bibitem{PLB}T.C. Kraan and P. van Baal, {\it Exact T-duality between calorons
and Taub-NUT spaces}, Phys. Lett. {\bf B428} (1998) 268 (hep-th/9802049).
\bibitem{Taubes}C. Taubes, {\em Morse theory and monopoles: topology in 
long range forces}, in: {\it Progress in gauge field theory}, eds. G. 't 
Hooft et al, (Plenum Press, New York, 1984) p. 563.
\bibitem{LeeLu}K. Lee, {\it Instantons and magnetic monopoles on $R^3 \times 
S^1$ with arbitrary simple gauge groups}, hep-th/9802012; K. Lee and C. Lu, 
{\it SU(2) calorons and magnetic monopoles}, hep-th/9802108.
\bibitem{HarShe}B.J. Harrington and H.K. Shepard, Phys. Rev. {\bf D17} (1978) 
2122; ibid. {\bf D18} (1978) 2990.
\bibitem{GroPisYaf}D.J. Gross, R.D. Pisarski and L.G. Yaffe, Rev. Mod. Phys. 
{\bf 53} (1983) 43.
\bibitem{BPr}P. van Baal, {\it Twisted boundary conditions: a non-perturbative
probe for pure non-abelian gauge theories}, (Ph.D. thesis, Utrecht, July 1984).
\bibitem{NahFou}W. Nahm, Phys. Lett. {\bf 90B} (1980) 413.
\bibitem{NahAll}W. Nahm, {\it All self-dual multimonopoles for arbitrary 
gauge groups}, CERN preprint TH-3172 (1981), published in Freiburg ASI
301 (1981); {\it The construction of all self-dual multimonopoles by the ADHM 
method}, in: ``Monopoles in quantum field theory", eds. N. Craigie, e.a. 
(World Scientific, Singapore, 1982), p.87.
\bibitem{Abproj}G. 't Hooft, Nucl. Phys. {\bf B190[FS3]} (1981) 455; Physica 
Scripta   {\bf 25} (1982) 133.
\bibitem{BraBaa}P.J. Braam and P. van Baal, Commun. Math. Phys. {\bf 122} 
(1989) 267.
\bibitem{BaaT3R}P. van Baal, Nucl. Phys. {\bf B}(Proc. Suppl.) {\bf 49} (1996) 
238 (hep-th/9512223).
\bibitem{DonKro}S.K. Donaldson and P.B. Kronheimer, {\it The Geometry of 
Four-Manifolds}, (Clarendon Press, Oxford, 1990).
\bibitem{AtiSinInd}M.F. Atiyah, I.M. Singer, Ann. Math. {\bf 93} (1971) 119.
\bibitem{CalBotSee}C. Callias, Commun. Math. Phys. {\bf 62} (1978) 213;
R. Bott and R. Seeley, Commun. Math. Phys. {\bf 62} (1978) 235.
\bibitem{Hoo}G. 't Hooft, Phys. Rev. {\bf D14} (1976) 3432.
\bibitem{CorGod}E. Corrigan and P. Goddard, Ann. Phys. (N.Y.) {\bf 154} 
(1984) 253.
\bibitem{Hit}N.J. Hitchin, Commun. Math. Phys. {\bf 89} (1983) 145;
J. Hurtubise and M.K. Murray, Commun. Math. Phys. {\bf 122} (1989) 35.
\bibitem{Nak}H. Nakajima, {\it Monopoles and Nahm's Equations}, in ``Einstein 
metrics and Yang-Mills connections", Sanda, 1990, eds. T. Mabuchi and S. Mukai,
(Dekker, 1993, New York).
\bibitem{AtiFerLec} M.F. Atiyah, {\it Geometry of Yang-Mills fields}, Fermi 
lectures, (Scuola Normale Superiore, Pisa, 1979).
\bibitem{Temp}E.F. Corrigan, D.B. Fairlie, S. Templeton and P. Goddard,
Nucl. Phys. {\bf B140} (1978) 31.
\bibitem{JacNohReb}G. 't Hooft, unpublished, quoted in R. Jackiw, C. Nohl, 
C. Rebbi, Phys. Rev. {\bf D15} (1977) 1642; R. Rajaraman, {\it Solitons and 
Instantons}, (North-Holland, Amsterdam, 1982).
\bibitem{Osb}H. Osborn, Ann. Phys. (N.Y.) {\bf 135} (1981) 373.
\bibitem{Kraan}T.C. Kraan, {\it Nahm's transformation and instantons on 
$R^3 \times S^1$}, poster presented at the 2nd DRSTP symposium, Dalfsen, 
The Netherlands, 5-6 June, 1997 (unpublished).
\bibitem{WhiWat}E.T. Whitakker and G.N. Watson, {\it A course in modern
analysis}, (Cambridge Univ. Press, 1927) p. 169. 
\bibitem{Ros}P. Rossi, Nucl. Phys. {\bf B149} (1979) 170.
\bibitem{AtiHit} M.F. Atiyah and N.J. Hitchin, {\it The Geometry and Dynamics 
of Magnetic Monopoles}, (Princeton Univ. Press, 1988).
\bibitem{Nut} E.T. Newman, T. Unti and L. Tamburino, J. Math. Phys. {\bf 4}
(1963) 915.
\bibitem{Haw1}S. Hawking, in: {\it General Relativity, an Einstein centenary 
survey}, eds. S. Hawking and W. Israel, (Cambridge Univ. Press, 1979) p. 774.
\bibitem{Haw2}S. Hawking, Phys. Lett {\bf 60A} (1977) 81.
\bibitem{SdEi}T. Eguchi and A.T. Hanson, Phys. Lett. {\bf 74B} (1978) 249;
G. Gibbons and C. Pope, Commun. Math. Phys. {\bf 66} (1979) 267.
\bibitem{HKLR}N.J. Hitchin, A. Karlhede, U. Lindstr\"om and M. Ro\v cek,
Commun. Math. Phys. {\bf 108} (1987) 535.
\bibitem{GauHar}J.P. Gauntlett and J.A. Harvey, {\it S-Duality and the 
Spectrum of Magnetic Monopoles in Heterotic String Theory}, hep-th/9407111.
\bibitem{SoGrPe} R. Sorkin, Phys. Rev. Lett. {\bf 51} (1983) 87;
D. Gross and M.Perry, Nucl. Phys. {\bf B226} (1983) 29.
\bibitem{GibMan}G.W. Gibbons and N.S. Manton, Nucl. Phys. {\bf B274} (1986) 
183.
\bibitem{Man}N.S. Manton, Phys. Lett. {\bf 154B} (1985) 397 [Err. {\bf 157B}
(1985) 475].
\bibitem{Con} S.A. Connell, {\it The dynamics of the SU(3) charge (1,1) 
magnetic monopole} (1991), ftp://maths.adelaide.edu.au/pure/murray/oneone.tex,
unpublished preprint.
\bibitem{su3}J.P. Gauntlett and D.A. Lowe, Nucl. Phys. {\bf B472} (1996) 194 
(hep-th/9601085); K. Lee, E.J. Weinberg and P. Yi, Phys. Lett. {\bf B376} 
(1996) 97 (hep-th/9601097).
\bibitem{LeeYi}K. Lee and P. Yi, {\it Dyons in $N=4$ supersymmetric theories
and three-pronged strings}, hep-th/9804174.
\bibitem{Tau}C. Taubes, J. Diff. Geom. 19 (1984) 517.
\bibitem{TauCMP}C. Taubes, Commun. Math. Phys. {\bf 86} (1982) 257, 299.
\bibitem{Mon}G. 't Hooft, Nucl. Phys. {\bf B79} (1974) 276; A.M. Polyakov,
JETP Lett. {\bf 20} (1974) 194.
\bibitem{JulZee}B. Julia and A. Zee, Phys. Rev. {\bf D11} (1975) 2227.
\bibitem{Wit}E. Witten, Phys. Lett. {\bf 86B} (1979) 283.
\bibitem{Manton}N. Manton, Phys. Lett. {\bf 110B} (1982) 54.
\bibitem{Super}G. 't Hooft, in: ``High Energy Physics", EPS conference, 
Palermo, 1975, ed. A. Zichichi (Editrice Compositori, Bologna, 1976); 
S. Mandelstam, Phys. Rept. {\bf 23C} (1976) 245
\bibitem{Shur}E. Shuryak, Phys. Rep. {\bf 264} (1996) 357; T. Sch\"afer and
E. Shuryak, Rev. Mod. Phys. {\bf 70} (1998) 323 (hep-ph/9610451).
\bibitem{Smit}J. Smit and A. van der Sijs, Nucl. Phys. {\bf B355} (1991) 603.
\bibitem{MaxAb}A.S. Kronfeld, G. Schierholz and U.J. Wiese, Nucl. Phys. 
{\bf B293} (1987) 461.
\bibitem{MonDom}T. Suzuki and I. Yotsuyanagi, Phys.Rev. {\bf D42} (1990) 4257;
M. Polikarpov, Nucl.Phys. {\bf B}(Proc. Suppl.){\bf 53} (1997) 134 
(hep-lat/9609020), and ref. therein. 
\bibitem{MonInst}M.N. Chernodub and F.V. Gubarev, JETP Lett. {\bf 62} (1995)
100 (hep-th/9506026); A. Hart and M. Teper, Phys.Lett. B372(1996) 261
(hep-lat/9511016); V. Bornyakov and G. Schierholz, Phys. Lett. B384(1996)190
hep-lat/9605019); R.C. Brower, K.N. Orginos and Ch.-I. Tan, Phys. Rev. {\bf D55}
(1997) 6313 (hep-th/9610101)
\bibitem{Edin}P. van Baal, Nucl. Phys. {\bf B}(Proc. Suppl.) {\bf 63A-C} (1998) 
126 (hep-lat/9709066).
\bibitem{ICool}M. Garc\'{\i}a P\'erez, A. Gonz\'alez-Arroyo, J. Snippe and 
P. van Baal, Nucl.Phys. B413(1994)535-553 (hep-lat/9309009).
\bibitem{LauSch}M.L. Laursen and G. Schierholz, Z. Phys. {\bf C38} (1988) 501.
\end{thebibliography}
\end{document}